\documentclass[amsmath,amssymb,amsfonts,twocolumn,pre]{revtex4-2}

\pdfoutput=1

\usepackage{graphicx}

\usepackage[colorlinks, citecolor=blue]{hyperref}

\usepackage{xspace}
\usepackage{subfigure}
\usepackage[utf8]{inputenc}
\usepackage[english]{babel}
\usepackage{txfonts}
\usepackage{natbib}

\newcommand{\citeg}[1]{\cite{#1}}

\usepackage{booktabs}

\usepackage{hyperref}
\newcommand{\safeurl}[1]{\href{#1}{#1}}

\usepackage{mathrsfs}

\newcommand{\nn}{\nonumber}
\providecommand{\E}[1]{\ensuremath{\mathbb{E}\left[#1\right]}}

\usepackage{cleveref}
\usepackage{bm}

\newcommand{\refeq}[1]{\cref{eq:#1}}
\newcommand{\sect}[1]{sect.~\ref{sec:#1}}
\newcommand{\Fig}[1]{Fig.\ref{fig:#1}}
\newcommand{\lgg}[1]{\ensuremath{\mathcal{L}_\alpha(#1)}}
\newcommand{\N}[1]{\ensuremath{\mathcal{N}(#1)}}
\newcommand{\mean}[1]{\ensuremath{\langle#1\rangle}}
\newcommand{\kmean}{\mean{k}\xspace}
\providecommand{\norm}[1]{\lVert#1\rVert}
\providecommand{\abs}[1]{\lvert#1\rvert}

\newcommand{\Nsim}{\ensuremath{N_\mathrm{sim}\xspace}}
\newcommand{\Nclus}{\ensuremath{N_\text{clus}\xspace}}

\begin{document}
\title{Levy geometric graphs}
\author{S. Plaszczynski}
\email{stephane.plaszczynski@ijclab.in2p3.fr}
\author {G. Nakamura}
\altaffiliation[Also at ]{RIKEN iTHEMS, Wako, Saitama 351-0198, Japan}
\author{C. Deroulers}
\author{B. Grammaticos}
 \author{M. Badoual}
\affiliation{ Universit\'e Paris-Saclay, CNRS/IN2P3, IJCLab, 91405
  Orsay, France}
\affiliation{Universit\'e Paris-Cit\'e, IJCLab, 91405 Orsay France}
\date{\today}
\begin{abstract}
We present a new family of graphs with  remarkable
properties. They are obtained by connecting the points of a 
random walk when their distance is smaller than a given scale.
Their degree (number of neighbors) does not depend on the graph's size but only on the
considered scale. It follows a Gamma distribution and thus presents an
exponential decay.
Levy flights are particular random walks with some power-law
increments of infinite variance. When building the geometric graphs
from them, 
we show from dimensional arguments, that the number of connected
  components (clusters) follows an inverse power of the scale.
The distribution of the size of their components, properly
normalized, is scale-invariant, 
which reflects the self-similar nature of the underlying process.
This allows to test if a graph (including non-spatial ones) could possibly result
 from an underlying Levy process.
When the scale increases, these graphs never tend towards a single
cluster, the giant component.
In other words, while the autocorrelation of the process scales as a
power of the distance,
they never undergo a phase transition of percolation
type.
The Levy graphs may find applications in community detection and in 
the analysis of collective behaviors as in face-to-face
interaction networks.
\end{abstract}

\maketitle
 

\section*{Introduction}
Graphs describe \textit{a set of relations} (edges) among
some \textit{objects} (vertices) and are thus the fundamental 
entities for analyzing interactions in complex systems.
The celebrated work of 
Erd\"os and R\'enyi \cite{ER:1959,ER:1960,ER:1961} 
marks the beginning
of graph structure exploration. In this reference model, still often used
today to generate null tests, edges are
randomly chosen among all possibilities with some given probability $p$. 
Many results have been
established for these graphs, the most salient feature being that a
transition similar to percolation\cite{AB:2002} appears beyond some critical
connectivity ($p_c=1/N$, $N$ being the graph size) with a ``giant
component'' containing an extensive number of vertices.
The distribution of the number of neighbors in these graphs (called
the degree) is a Poisson one and therefore strongly peaked around the mean
value $pN$ especially when this one is large.

Graphs embedded in space, i.e. where each vertex has some
  associated coordinates, are often called ``spatial networks''.
Some adaptation of random graphs to them was proposed
\cite{Gilbert:1961} by linking nearby points. The resulting graph is now called 
a \textit{random geometric graph} (RGG). The standard procedure
is to first populate randomly $N$  points in the plane
and creates the edge $e_{ji}$ if the distance between the $i$ and $j$
vertices is below some given cutoff $d(i,j)<R$.
The resulting geometric graph is closely related to a pure random one with a
connection probability $p=\pi R^2$ (in dimension 2, assuming a unit
total surface) and a mean degree
\begin{align}
\label{eq:meandegRGG}
  \kmean=p N=\pi R^2 N.
\end{align}
RGGs also exhibit a critical transition above which a giant component
develops which happens around $\kmean_c=4.5$ in two dimensions (2D) \cite{Dall:2002}.
Although there exist some differences between pure random graphs and
geometric ones, in particular on the density of triangles, 
the degree distribution of RGGs is still a Poisson one while many
real-world networks are more heavy-tailed, going up to power-law
(scale-free) distributions \cite{Newman:2003b}. 
In spatial networks, cost considerations (energetical, economical) tend to restrict the
appearance of very large degrees \cite{Barthelemy:2011}, but the
degree distributions are still broad.

Several works have focused on ways to obtain a scale-free degree
distribution. For RGG, this can be achieved by
changing the probability distribution of the points from uniform to a
more general form $p(\bm{x})$ \cite{Herrmann:2003}, or by changing 
the space geometry to a hyperbolic one \cite{Krioukov:2010}.
But the most influential step in that direction is the one by Barab{\'a}si and Albert \cite{BA:1999}
who introduced the notion of
\textit{growth} (one starts with very few vertices and then adds new
ones) and \textit{preferential attachment} (edges are connected
depending on the degree of the already present vertices).
The success of this approach  somewhat shifted 
the paradigm for graph generation and representation \cite{AB:2002} to
an iterative process governed by some rules, 
tightening the links with statistical physics. 

Random walks have a long and rich history \cite{Kampen:2007,Hugues:1995} and are of capital importance in statistical physics. 
By random-walk we loosely speak about the repeated sum of the same stochastic
processes (steps) and we will restrict ourselves to continuous
processes in space. 
The standard one is based on normally distributed increments (Wiener process)
and most walks converge to it 
since the sum of random variables always converges to a
Gaussian thanks to the Central Limit Theorem. This is in fact only
valid if the variance of the increment is finite. More generally, the
generalized central limit theorem \cite{Durrett:2010} states that the sum of any distribution, even
with an infinite variance, converges to a stable distribution for
which the normal distribution is a particular case.

In what follows we wish to connect the two domains of graph structure exploration
and stochastic processes by building a geometric graph from random-walk points.
Since power-law interactions are ubiquitous in physics and biology we
will put particular emphasis on Levy flights which lead to some remarkable
graph properties.

We will first review in \sect{lgg} the fundamentals of Levy flights and
the type of geometric graphs produced from them which we shall call Levy Geometric Graph
(LGG), generalizing them to any dimension and discussing the effect of
dimensionality. We will then discuss
in \sect{meandeg} the degree of the graph, making thus a first
connection with the random walk properties.
In \sect{clusters} we study the number and size of the connected components which have
some unique properties, and give insights about their structure.
Finally, in \sect{rw}, we shall compare these results to the ones obtained
with standard (Gaussian) random walks that will help understand what
makes the Levy graphs special. We shall conclude with some possible
applications, and defer to more technical Appendices the computation of the
autocorrelation function for a 2D Levy process and of the mean
degree of a standard random walk graph.

\section{Construction}
\label{sec:lgg}

\subsection{Levy flight}
\label{sec:levy}

\begin{figure}
  \subfigure[]{\includegraphics[width=8.20cm]{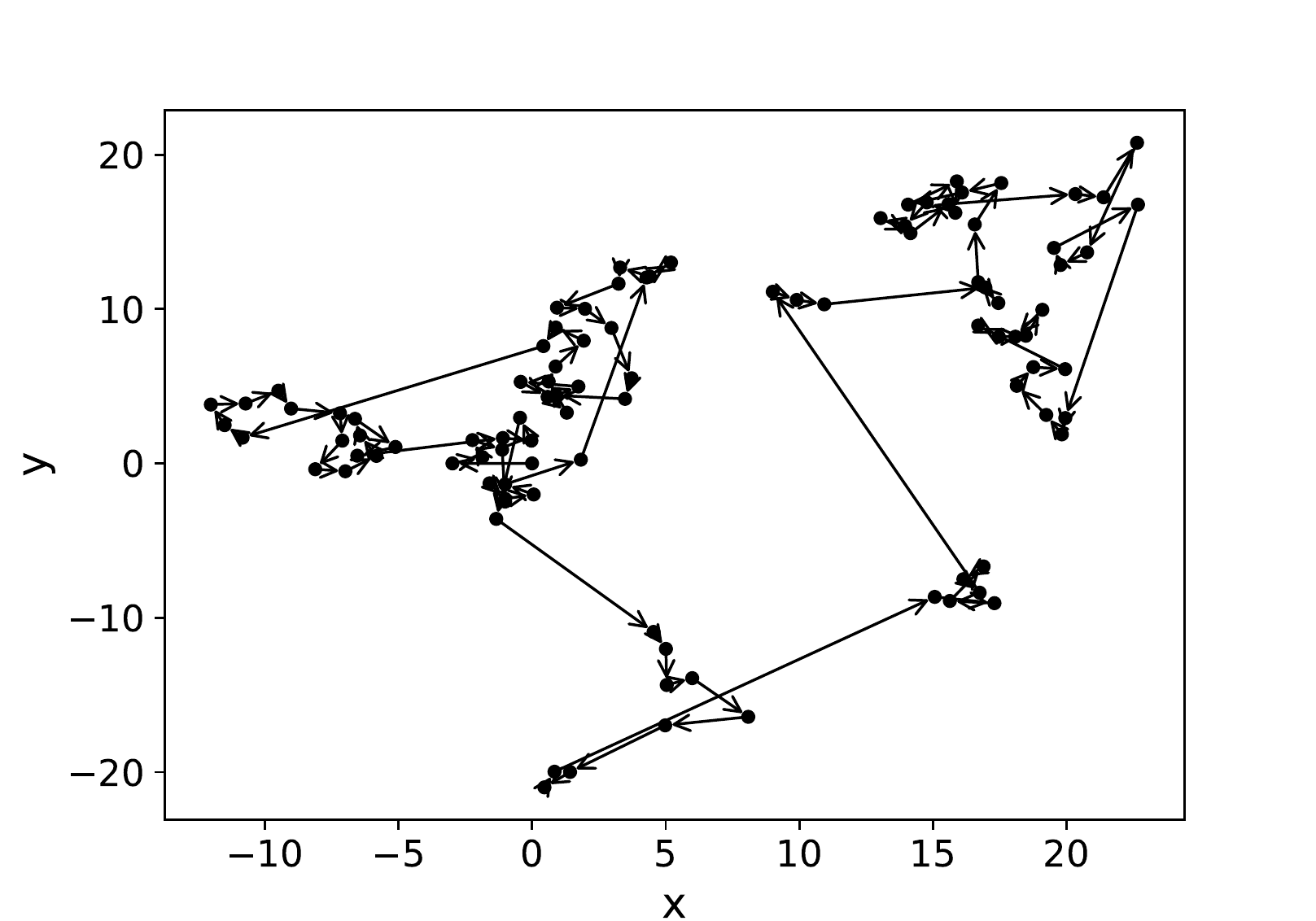}}\\
  \subfigure[s=3]{\includegraphics[width=8.20cm]{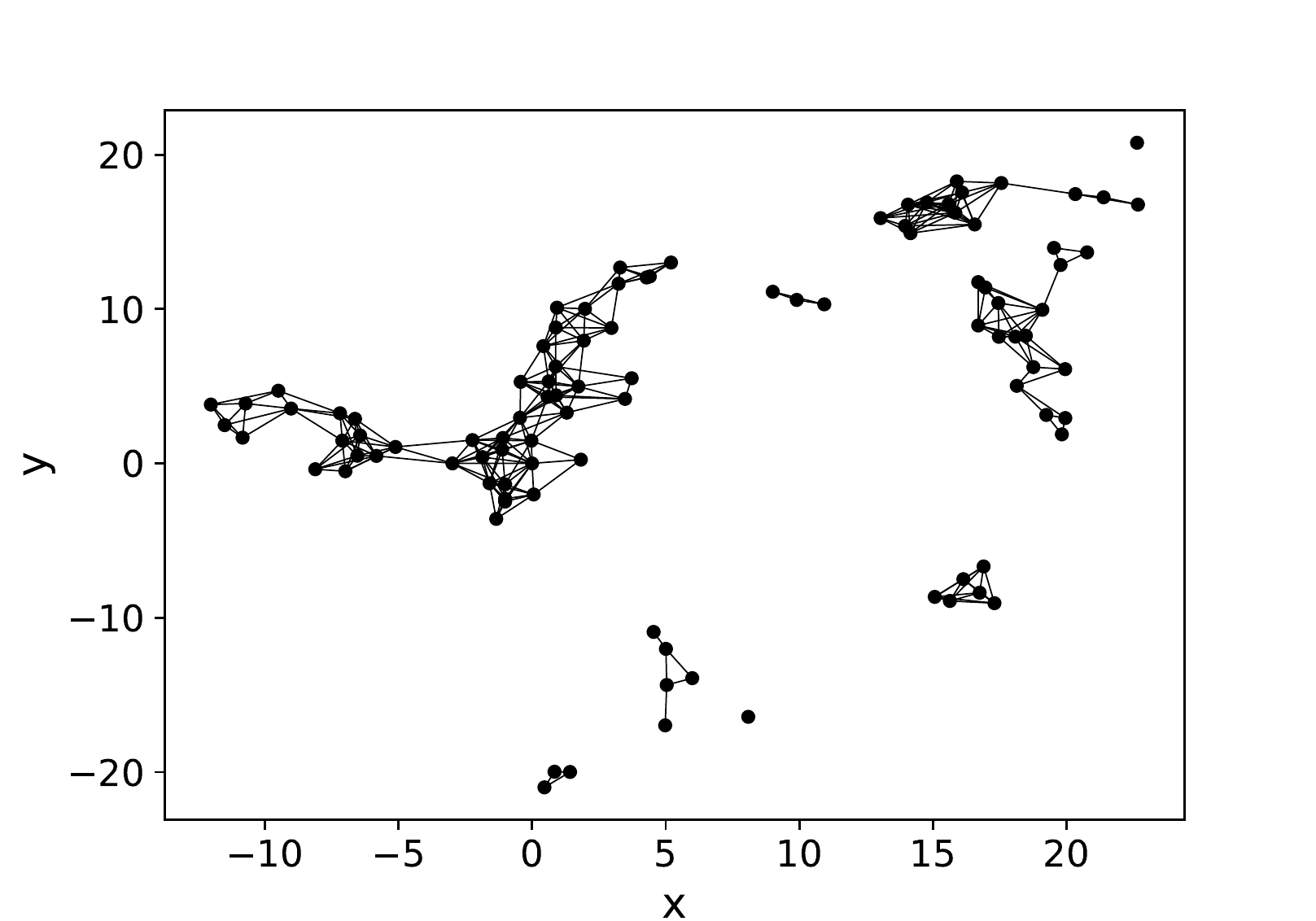}}\\
  \subfigure[s=5]{\includegraphics[width=8.16cm]{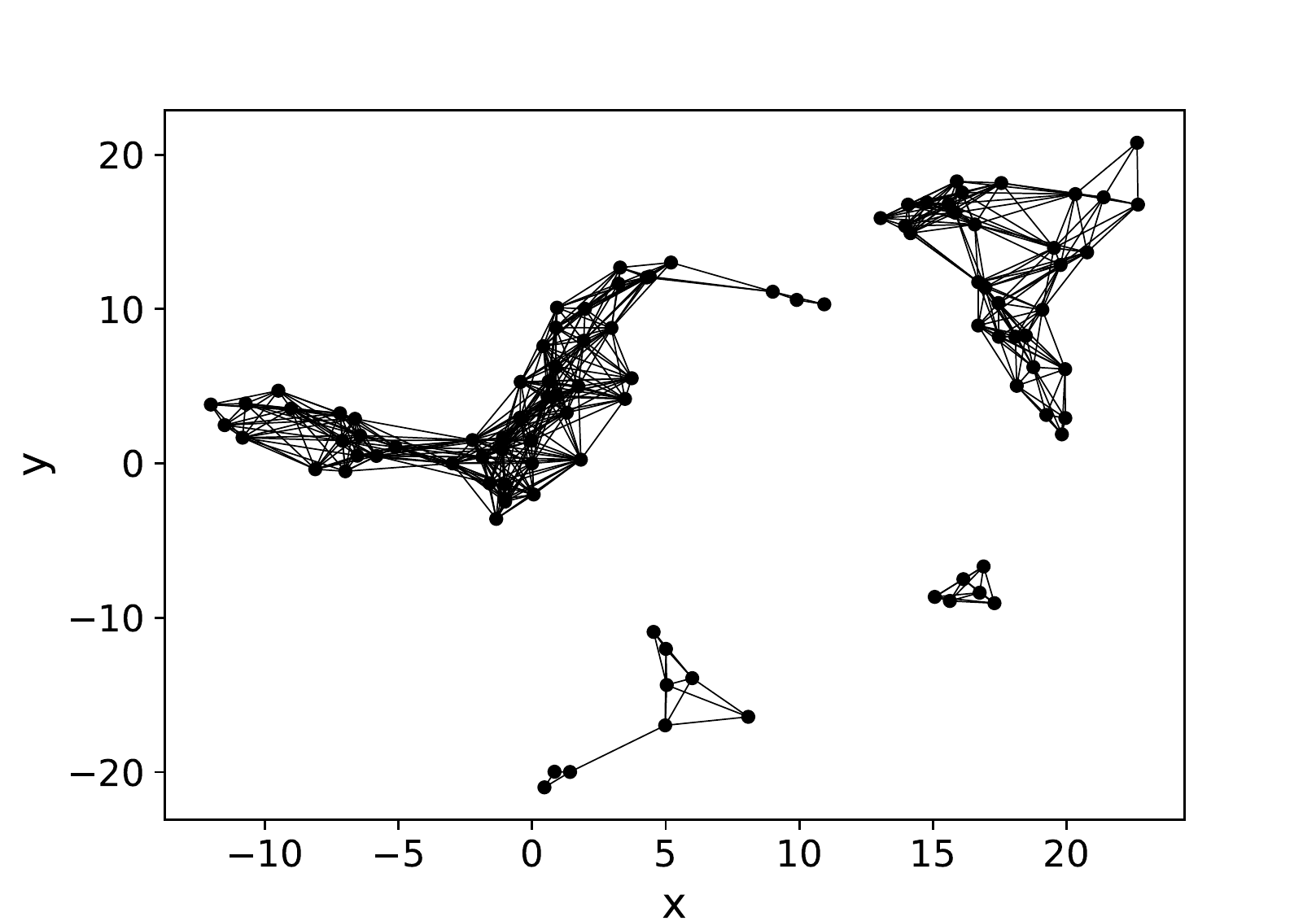}}
  \caption{\label{fig:lgg} (a) Example of a Levy flight
    ($\alpha=1.5,r_0=1, N=100$) and (b,c) of two geometric graphs built from
    it at different scales}
\end{figure}

Mandelbrot\cite{Mandelbrot:1975,Mandelbrot:1983} has introduced the concept of Levy flight
(or walk) as a tribute to his teacher's work on stable distributions
(for an introduction, see \cite{Chechkin:2006}). The method consists
first in drawing some radial random number ($X$) according to a power-law
distribution but only above some cutoff value ($r_0$). Mandelbrot dubbed it the
Pareto-Levy distribution. Its cumulative distribution
  function (also called survival probability)
  is
\begin{align}
\label{eq:surv}
P(X>r)=
  \begin{cases}
    \left(\dfrac{r_0}{r}\right)^\alpha  & \text{for}~ r \geq r_0 \\
    1  & \rm{else},
  \end{cases}
\end{align}
which, by taking the derivative, gives for the probability density function
\begin{align}
\label{eq:f1}
  f(r)=
  \begin{cases}
      \dfrac{\alpha}{r_0}\left(\dfrac{r_0}{r}\right)^{1+\alpha}~
      &\text{for}~r \geq r_0 \\
    0 &\text{for}~ r <r_0.
\end{cases}
\end{align}
An interesting feature of this distribution is that for the Levy index $\alpha<2$ its
variance is infinite, meaning that for samples drawn according
to it, the measured standard deviation does not converge with the
sample size. 
From \refeq{surv} one derives a straightforward way of drawing
numbers according to a Pareto-Levy distribution by first drawing a value $u_i$ from a
$[0,1]$ uniform distribution and transforming it according to $r_0 u_i^{-1/\alpha}$.
By also drawing an isotropic angle in $[0,2\pi]$, we obtain the coordinates of a point
and build the random walk by accumulating the Euclidean positions 
(see an example in \Fig{lgg}(a)).

The properties of such a random point process are very unusual to
scientists familiar with the convergence properties coming from the
Central Limit Theorem, which is not applicable here due to infinite variances. 
The process is actually \textit{non}-homogeneous; there is no
mean density as in a Poisson process, or, in the point-process vocabulary \cite{Cox:1980}, a
first-order intensity function. However the process  
has an isotropic autocorrelation function (second order intensity function) defined as the
\textit{conditional} probability of finding a point at a distance $r$
from a point of the process. Its computation is explained in
  detail in Appendix
\ref{app:xi} and leads to
\begin{align}
\label{eq:autocorr}
  f(\bm r)\propto \dfrac{1}{r^{2-\alpha}},~\text{for}~0<\alpha<2~\text{and}~r\gg r_0.
\end{align}
This power-law behavior can be understood considering 
  the asymptotic tail of the Pareto-Levy distribution (\refeq{f1})
  which is that of a stable distributions \cite{Chechkin:2008} with
  a characteristic function $\phi(k)=
  e^{-ck^\alpha}\simeq 1-ck^\alpha$ in the low $k$ (large $r$) limit. Fourier-integrating it on a space of dimension 2 leads to the result.

By integrating the process on a disk of radius $R$, one then finds that the mean
number of points in it is
\begin{align}
  \label{eq:mandel}
  \bar N(<R)\propto\left(\dfrac{R}{r_0}\right)^\alpha,~\text{for}~0<\alpha<2~\text{and}~R\gg r_0,
\end{align}
exhibiting a fractal dimension in the power law. 
A process with a power-law autocorrelation function is scale-free
or more precisely self-similar \cite{Newman:2005}.

We emphasize that these results rely on some approximations
that we highlight in Appendix \ref{app:xi}.
In particular it is sometimes stated that for $\alpha \ge 2$ the
process becomes Gaussian. Although this is valid for large
values of $\alpha$ we show that  
this transition is progressive. While the power-law description is
excellent for $\alpha$ values close to 1, around $\alpha=2$ the conditional
distribution becomes
a complicated mixture of power-law and Gaussian functions.

\subsection{Levy Geometric Graphs (LGG)}
\label{sec:graph}

The Levy flight is an oriented path.
We obtain an undirected graph by applying some scale, i.e. we
consider it at some given ``resolution". We use the standard
geometric graph recipe by connecting points if their Euclidean distance
is below some cutoff value $R$:
\begin{align}
\label{eq:dij}
  \norm{\bm{X}_i-\bm{X}_j}\leq R.
\end{align}

What matters here is the relative value between the $R$ cutoff and the minimal
step size $r_0$ of the
Pareto-Levy distribution, so that, in what follows, we will only use the \textit{scale}
$ s\equiv\dfrac{R}{r_0}$ or, equivalently, always work setting $r_0$
to 1 so that $s$ represents the geometric cutoff.

Increasing the $s$ cutoff, one obtains fewer and fewer clusters which
become bigger and bigger as illustrated in \Fig{lgg} (b) and (c).
Although the resulting graph is a metric one (positions are properties of
the vertices) we will only consider their connectivity
structure.

For given $\alpha$ exponent and $s$ scale values, we
call the resulting graphs the Levy Geometric Graphs (LGG) and note them \lgg{s}.
The fractal properties are valid for $s\gg1$ (which will be made more
precise in \sect{meandeg}) and for $\alpha\leq2$. However
for $\alpha<1$ the mean of the Pareto-Levy distribution diverges and
all statistics are governed by rare events leading to very noisy
results. So we shall not consider $\alpha<1$ values.
In what follows, our range of interest for the LGG parameters will be
\begin{subequations}
\label{eq:range}
\begin{align}
  1&\leq\alpha\leq2  \\
  s&\geq 2
\end{align}  
\end{subequations}

\subsection{Dimensionality}
\label{sec:dim}

Although Levy flights are generally studied in dimension $d=$2 or 3
we generalize them to any other dimension $d$ by building the walk using
\refeq{surv} for the radius and drawing an isotropic direction, 
for instance from a standard $d$-dimensional normal distribution. The edge assignment is still
performed using \refeq{dij} in the $d$ dimensional space.

The conditional probability is similar to the 2D case by replacing the
exponent 2 in \refeq{autocorr} by $d$. The mean number of points in a ball of
radius $R$ (\refeq{mandel}) is then unchanged up to the normalization factor.

Levy flights may be viewed as a sequence of ``local" points
followed by some ``long" jump. Due to isotropy some new points may
``come back" close to some previous ones as in \Fig{lgg}. The
probability that this happens, that we call the ``return-probability",
should decrease with dimension, eventually 
going to 0 as $d\to\infty$ since the path will go to other parts of space. 

To be more quantitative, we define a return-probability for the Levy
process in the following way.
Let us first suppose that we have switched ``off'' the angular part of
the process and we only keep the radial steps in an additive way. Then all pairs of points are
separated by a distance of at least $r_0=1$. Building a Levy graph that
connects points \textit{below} $r_0=1$ (\lgg{s=1}) just leads to
a disconnected set of points where there are as many connected
components as points ($\Nclus=N$).
When switching the angular part ``on'', some points do come back close
to previous ones, sometimes below the $r_0=1$ cut, 
and some clusters start to form ($\Nclus<N$).
We then propose the following definition for a Levy flight return-probability
\begin{align}
  P_0=\lim_{N\to\infty}\left(1-\dfrac{\Nclus}{N}\right).
\end{align}
where \Nclus\ is the number of clusters in a \lgg{s=1} of size $N$.

We estimate those numbers in
dimensions 2 to 5 by building 100 \lgg{s=1} graphs ($N=10^5$), counting
each time the number of connected components, and computing
the mean and standard deviation of the $(1-\tfrac{Nclus}{N})$ values.
Results are reported in table \ref{tab:P0}.

\begin{table}
\centering
  \begin{ruledtabular}
\begin{tabular}{rrrr}
  $d$ & $\alpha=1$ & $\alpha=1.5$ & $\alpha=2$ \\
\midrule
 2 & 0.192$\pm$0.002 & 0.444$\pm$0.015 & 0.709$\pm$0.036 \\
 3 & 0.067$\pm$0.001 & 0.141$\pm$0.002 & 0.233$\pm$0.002 \\
 4 & 0.031$\pm$0.000 & 0.064$\pm$0.001 & 0.103$\pm$0.001 \\
 5 & 0.017$\pm$0.001 & 0.035$\pm$0.001 & 0.055$\pm$0.001 \\
\end{tabular}
  \end{ruledtabular}
  \caption{\label{tab:P0} Return-probability as defined in the text
    measured for Levy graphs with different Levy indices
  in several dimensions.
}
\end{table}

In dimension 2, the return-probability is between 19 and 71\% 
depending on the Levy index.
If we rescale the $d=3,4,5$ probabilities by $P_0(2)$ we obtain for $P_0(d=3,4,5)/P_0(2)$:
\begin{align}
  (0.350\pm0.006,0.163\pm0.003,0.089\pm0.003) & \quad\alpha=1,\nn\\
  (0.318\pm0.011,0.144\pm0.005,0.078\pm0.003) & \quad \alpha=1.5, \nn\\
  (0.329\pm0.017,0.145\pm0.008,0.078\pm0.004) & \quad \alpha=2, \nn
\end{align}
which shows in each case a strong effect between dimensions 2 and 3 (about a factor 3) ,
and then milder ones (about a factor 2) when going from dimensions 
$3\to4$ and $4\to5$. 
This effect essentially depends on the space dimension, not on the details of
the Levy walk ($\alpha$).
It's worth noticing that these values are similar to the ones
obtained for a standard random walk but on a lattice (i.e. a square
grid) where the relative probabilities (with respect to dimension 2) to come back to a previous site
are \cite{Montroll:1956}
\begin{align}
\label{eq:polya}
P(d)/P(2)=(0.340,0.193,0.135). 
\end{align}

\section{Degree}
\label{sec:meandeg}

We first consider the average degree of the graph.
For a geometric graph cut at some distance $R$, the number of
neighbors (degree) at a given vertex is the number of points within
a disk of radius $R$ centered on it minus one (the vertex itself). The mean degree is
then
\begin{align}
  \kmean=\bar N(<R)-1
\end{align}

From \refeq{mandel} we then use the following model for the mean degree
\begin{align}
\label{eq:kmean}
\kmean(s)=A_D s^{\alpha_D}-1
\end{align}
where the amplitude $A_D$ and power exponent $\alpha_D$ will be
adjusted from the results of simulations.

\begin{figure}
  \includegraphics[width=8.2cm]{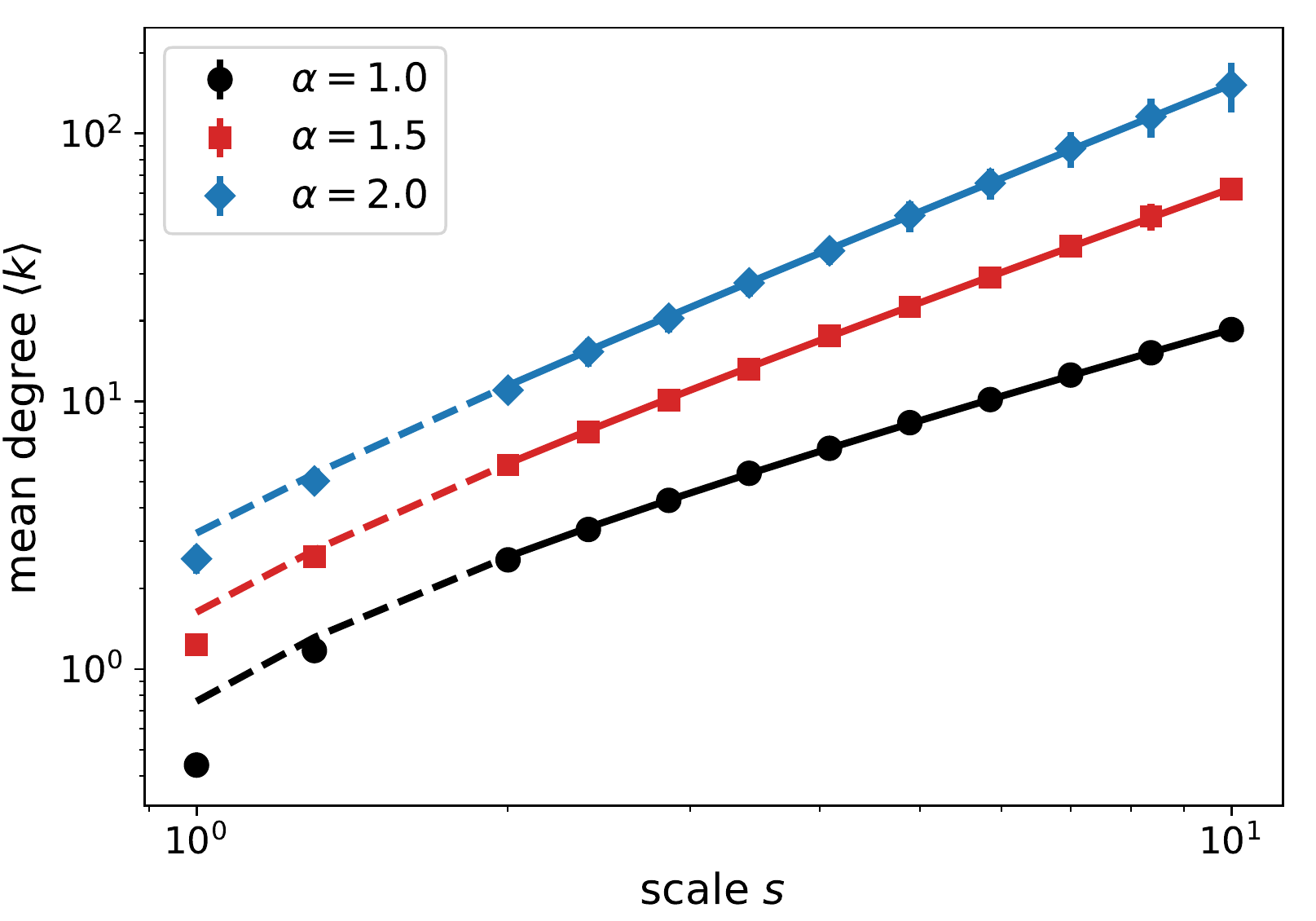}
  \caption{\label{fig:meandeg} Mean degree measured on dimension-2 Levy graphs
    varying the scale
    for several indices values. Full lines represent
    the \refeq{kmean} best fits
    performed in the $s\geq2$ region and the dashed ones their
    extension to lower values.
 }
\end{figure}

We measure the mean degree by running $100$ \lgg{s} simulations
of size $N=10000$ varying the scale and we show the average values with standard deviations
for $\alpha=1,1.5,2$ in \Fig{meandeg} together with the best fit to \refeq{kmean}.
The agreement is excellent down to $s=2$ which fixes our lower limit.
We have also checked that the power-law model agrees nicely for any
Levy index $\alpha$ and in any dimension.

\begin{figure}
  \includegraphics[width=8.16cm]{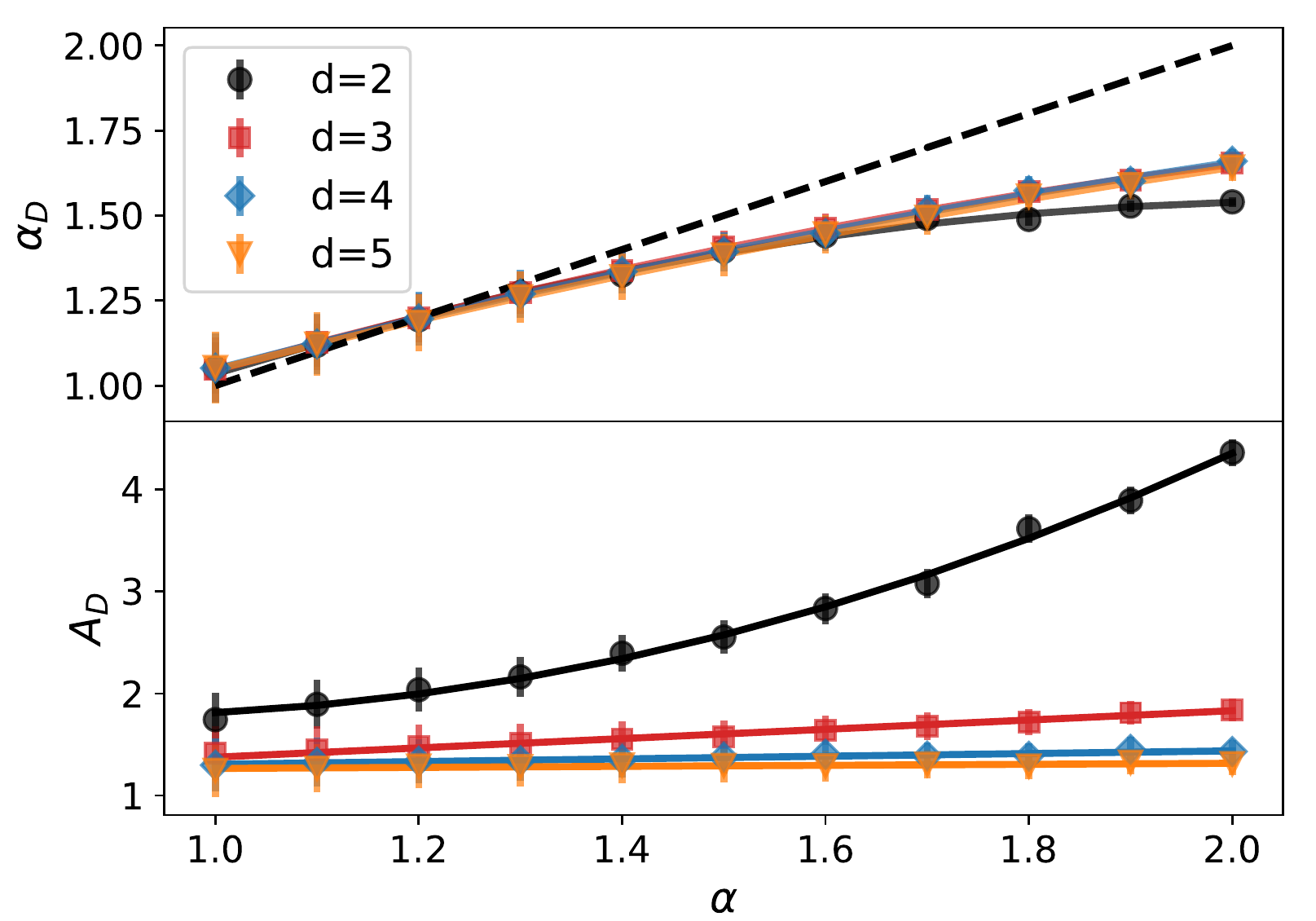}
  \caption{ Best fit parameters values measured on simulations from \lgg{s} mean degree according to the
    \refeq{kmean} model in several dimensions $d$. The points show
    the measured values and the lines the best quadratic fits (or linear
    in the case of
    $A_D$ for $d=3,4,5$).The upper dashed line
    shows the $\alpha_D=\alpha$ diagonal.}
  \label{fig:coeffs}
\end{figure}

\Fig{coeffs} shows the best-fit coefficients in several dimensions. 
For small values of $\alpha$, $\alpha_D\simeq\alpha $, but gets
  smaller when approaching 2. 
This is to be attributed to the 
approximations which entered in the derivation of \refeq{mandel} and that are
discussed in Appendix \ref{app:xi}. While $\alpha_D$ is practically independent of the dimension, the
amplitude parameter $A_D$ exhibits a strong dimension dependence. This
is due to the fact that, in low dimensions increasing the return-probability does increase the mean degree.

In dimension 2, one may use the following approximations:
  \begin{subequations}
  \label{eq:fitvals}
  \begin{align}
  \alpha_D(\alpha)&=\alpha-0.42(\alpha-1.21)(\alpha-0.60)\\
    A_D(\alpha)&=1.81+2.04(\alpha-1)(\alpha-0.75),
\end{align}
  \end{subequations}
and we note that the maximal value of $\alpha_D$ is around 1.5.

Finally we emphasize the following: 
\begin{itemize}
\item the mean degree fixes the total number of edges,
  $E=\kmean\tfrac{N}{2}$ for undirected graphs. Then for any \lgg{s}
  the mean number of edges is known. 
\item  the mean degree of a \lgg{s} is fixed by $\alpha$ and
  $s$ and is independent of the graph's size $N$.
\end{itemize}

\begin{figure}
  \includegraphics[width=8.16cm]{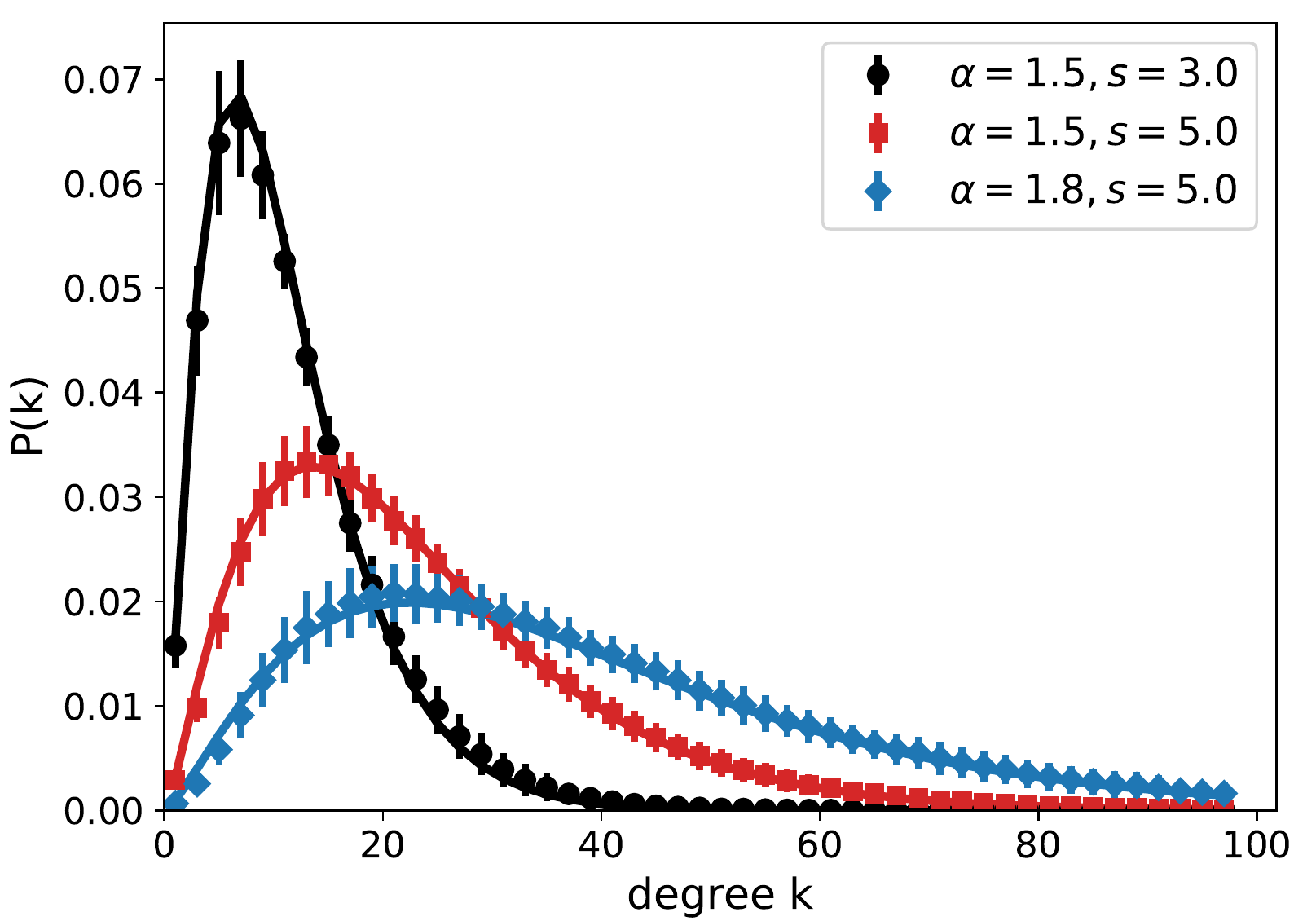}
  \caption{\label{fig:Pk_2d} Parametrization of the degree
    distribution in dimension 2, for some $(\alpha,s)$ values.
    The points with error bars show the mean of histograms built from
    100 simulations
    and the line, the analytical formula \refeq{anal} with $\beta=1.4$
}
\end{figure}

The  degree distribution has a tail because of points ``coming back''
to previous ones. We characterized it in \sect{dim} by a
return-probability, that only depends on the space dimension. 
We have noticed that in our range of parameters (\refeq{range}) the degree is
well described by a $\Gamma$ distribution 
 \begin{align}
\label{eq:anal}
   P(k)=\dfrac{k^\beta e^{-k/\theta} }{\Gamma(\beta+1)\theta^{\beta+1}},
\end{align}
where $\beta(d)$ depends on the dimension, and we set  
\begin{align}
   \theta=&\kmean/(\beta+1) \label{eq:theta}
\end{align}  
to ensure the proper mean value, since, for the $\Gamma$ distribution 
$\E{k}=\theta(\beta+1)=\kmean$. 
A fixed value of $\beta=1.4$ gives good fits for all
($\alpha,s$) values, as illustrated in \Fig{Pk_2d}.
Together with the mean degree formulas $\kmean(\alpha,s)$ 
  \cref{eq:kmean,eq:fitvals} , we then obtain an empirical 
parametrization of the degree distribution for any \lgg{s} (in
dimension 2). 
It shows that for large $k$ the tail decays essentially exponentially.

\section{Connected components}
\label{sec:clusters}

As is clear in \Fig{lgg}, the LGG construction
leads to a set of connected components (clusters) which are all
simple graphs.
Their number and sizes are random variables which we shall now characterize.

\subsection{Number of clusters}
\label{sec:Nclus}

\begin{figure}
  \includegraphics[width=8.16cm]{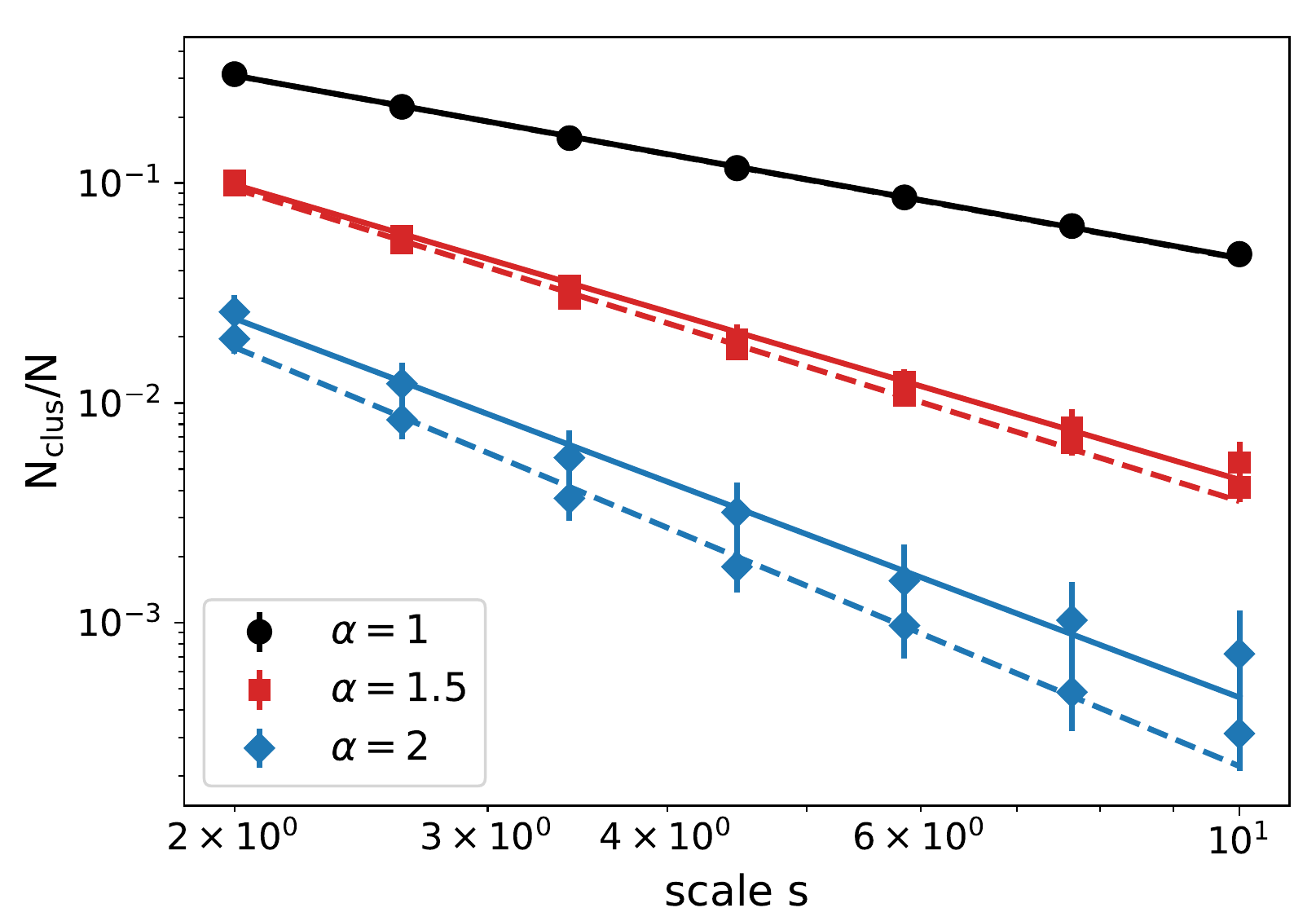}
  \caption{\label{fig:Nclus} Measured fraction of clusters (mean and standard
    deviation over 100 simulations at each point) for three LGGs varying
    the scale, for two graph's size.
    Full Lines show the the power-law model
    for $N=10^4$, and dashed ones for $N=10^5$. For
    $\alpha=1$ both are indistinguishable. The fitted exponents for
    $\alpha=(1,1.5,2)$ are respectively $(1.2,1.9,2.5)$ for $N=10^4$
    and $(1.2,2.0,2.7)$ for $N=10^5$.
    }
\end{figure}

We first look at the number of clusters as a function of the scale for
a given Levy index.
We measure it for two cases $N=10^4$ and $N=10^5$ on simulations
(\Nsim=100 for each point) by counting
the number of connected components.
\Fig{Nclus} shows the measured cluster fractions
for three $\alpha$ values varying the scale.  
They all follow a power-law function with similar slopes for
  the two $N$ values in particular when $\alpha\to1$. As for the mean degree case (\sect{meandeg}), the exponent is close to
$\alpha$ but here higher by about 25\%.

\begin{figure}
  \includegraphics[width=8.16cm]{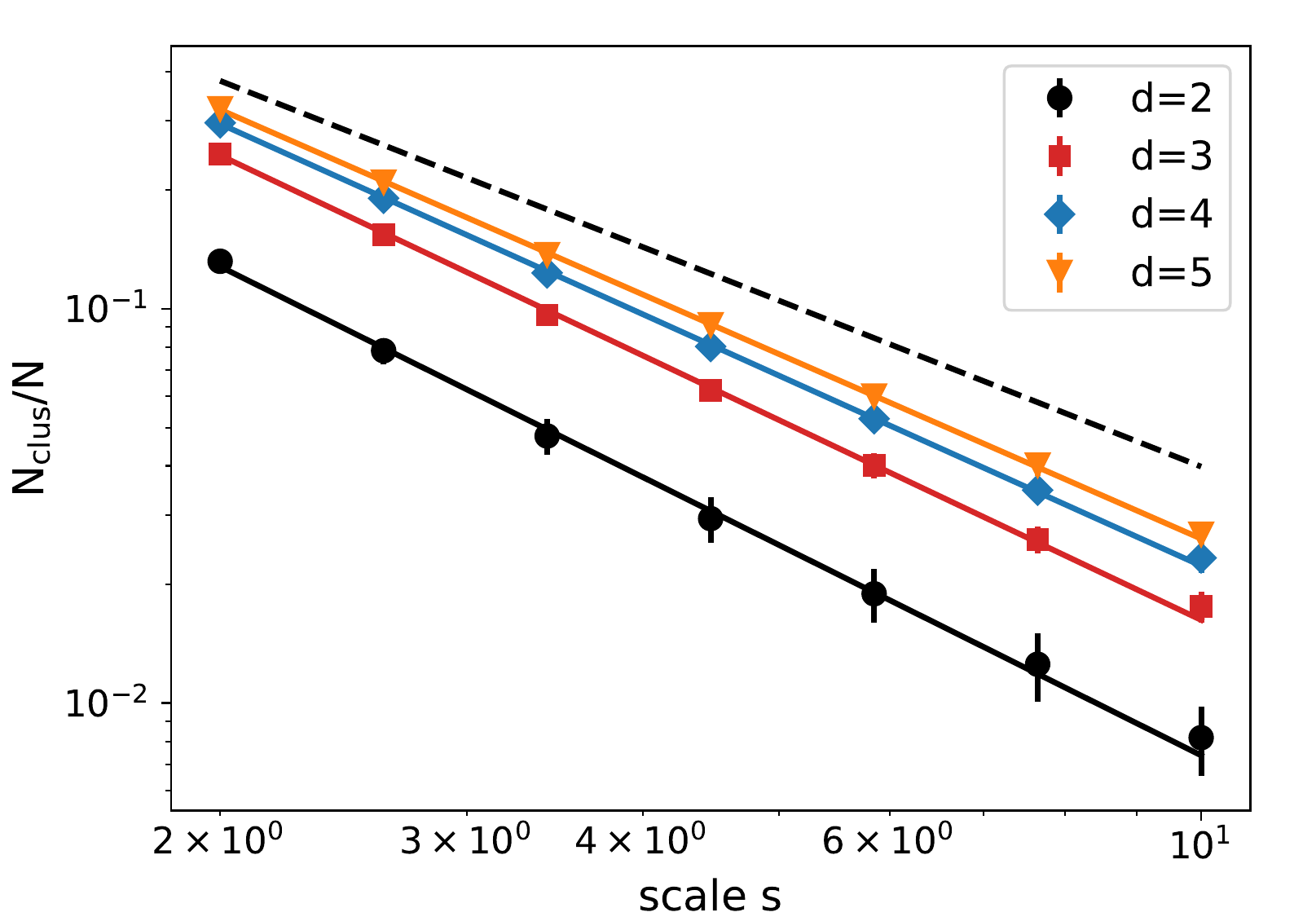}
  \caption{\label{fig:dims} Measured fraction of clusters of a Levy
    graph ($\alpha=1.4$) varying the space dimension $d\in[2,5]$ ($N=10^4$).
    The dashed line shows the asymptotic value $1/s^\alpha$ 
    reached for $d\to\infty$.
}
\end{figure}

To understand the origin of this scaling we may resort again to the
higher dimensional case where the return-probability 
may be neglected (\sect{dim}). In this case a cluster forms as soon as there is a
step larger than the $s$ scale. From \refeq{surv} this happens when
\begin{align}
  \label{eq:nclusexp}
  p(>s)=\dfrac{1}{s^\alpha}.
\end{align}
which shows the power dependency.
We show in \Fig{dims} how the cluster fraction varies when 
increasing the dimension. The cluster fraction
converges indeed to the \refeq{nclusexp} naive expectation 
following the pattern discussed in \sect{dim} (an important change
between dimensions 2 and 3 and then some milder ones). 
The logarithmic slope is unchanged, confirming the fact that the return-probability only affects the global normalization.

This also explains why the cluster fraction is mostly independent
of $N$. After a long jump, the  probability to have a
further one that brings back the walker near a previous point is very
small. Clusters are formed in different regions of space so that their number
scales about linearly with $N$.

\begin{figure}
  \includegraphics[width=8.16cm]{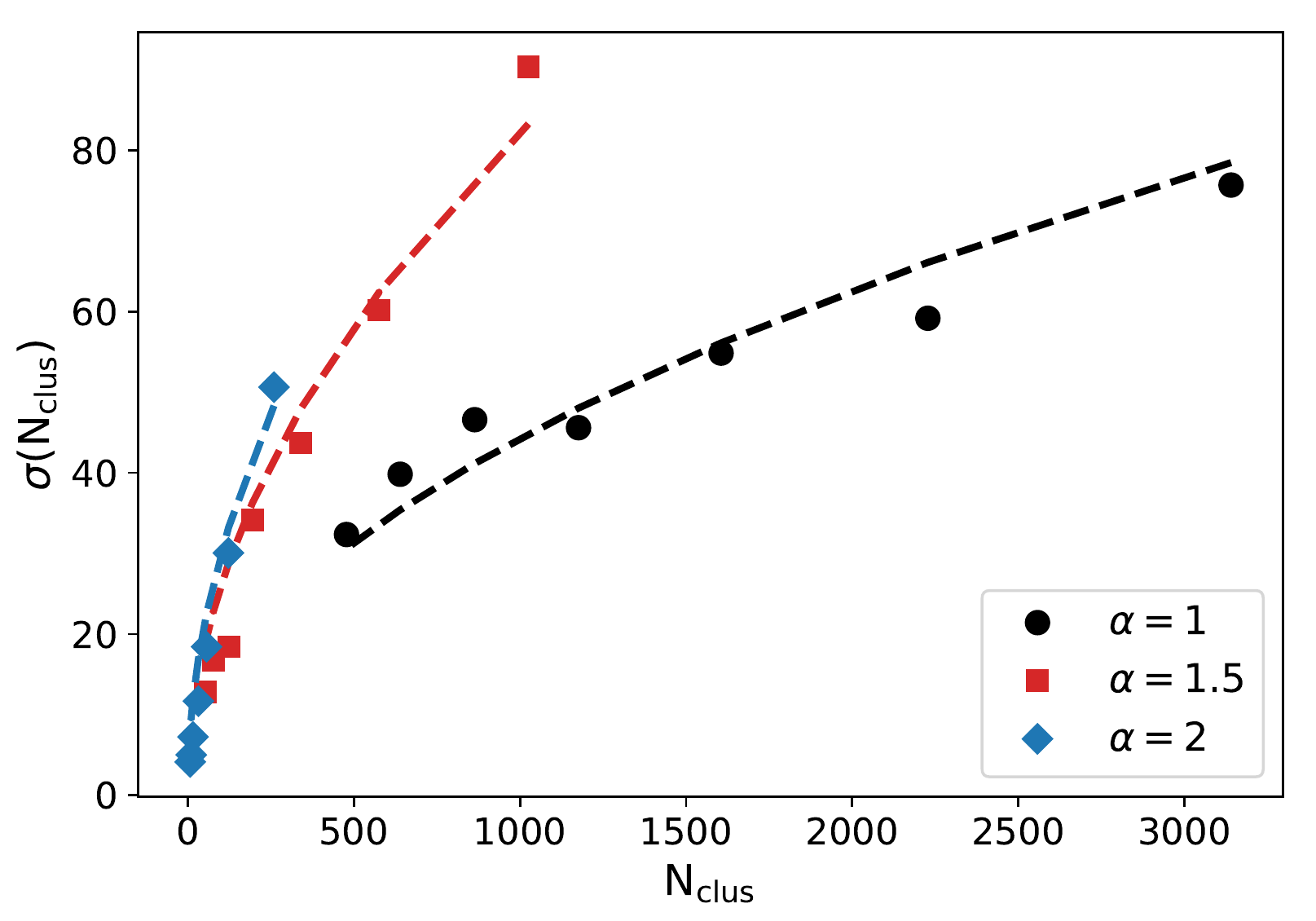}
  \caption{\label{fig:signclus} Standard deviation on the number of
    clusters as a function of their number for the three
    runs $\alpha=1,1.5$ and 2. The dashed lines show a square-root dependency.}
\end{figure}

It is also worth noticing that despite the fact that the process
is built from individual steps of infinite variance, the standard
deviation on 
the number of clusters is small.
We show in \Fig{signclus} that the standard deviation on the number of
clusters follows $\sigma(\Nclus)=b\sqrt{\Nclus}$ with $b=1.4,2.4,3$
for respectively $\alpha=1,1.5,2$. 
This is only a factor around two larger than for a Poisson process. 
This means that for any \lgg{s} graph, the number of
clusters in a run of length $N$ is \textit{a priori} known quite precisely.

\subsection{Cluster sizes}
\label{sec:clusize}
We now investigate the cluster sizes, i.e. the number of vertices of each connected  component.

For a LGG with $N$ vertices there are 
\Nclus\ clusters of various sizes $N_{i=1,...,\Nclus}$. Both $\Nclus$
and $N_i$s are the realization of random variables subject to the constraint
$N={\sum_{i=1}^{\Nclus} N_i}$. Obviously when there are
``fewer" clusters they should be ``larger" in order to preserve $N$.
In the following we weight the sizes by the  cluster fraction
and name it the \textit{normalized cluster size}:
\begin{align}
\label{eq:normsize}
  n_i=\dfrac{\Nclus}{N} \times N_i
\end{align}
and call $n$ the associated random variable.

\begin{figure}
  \includegraphics[width=8.16cm]{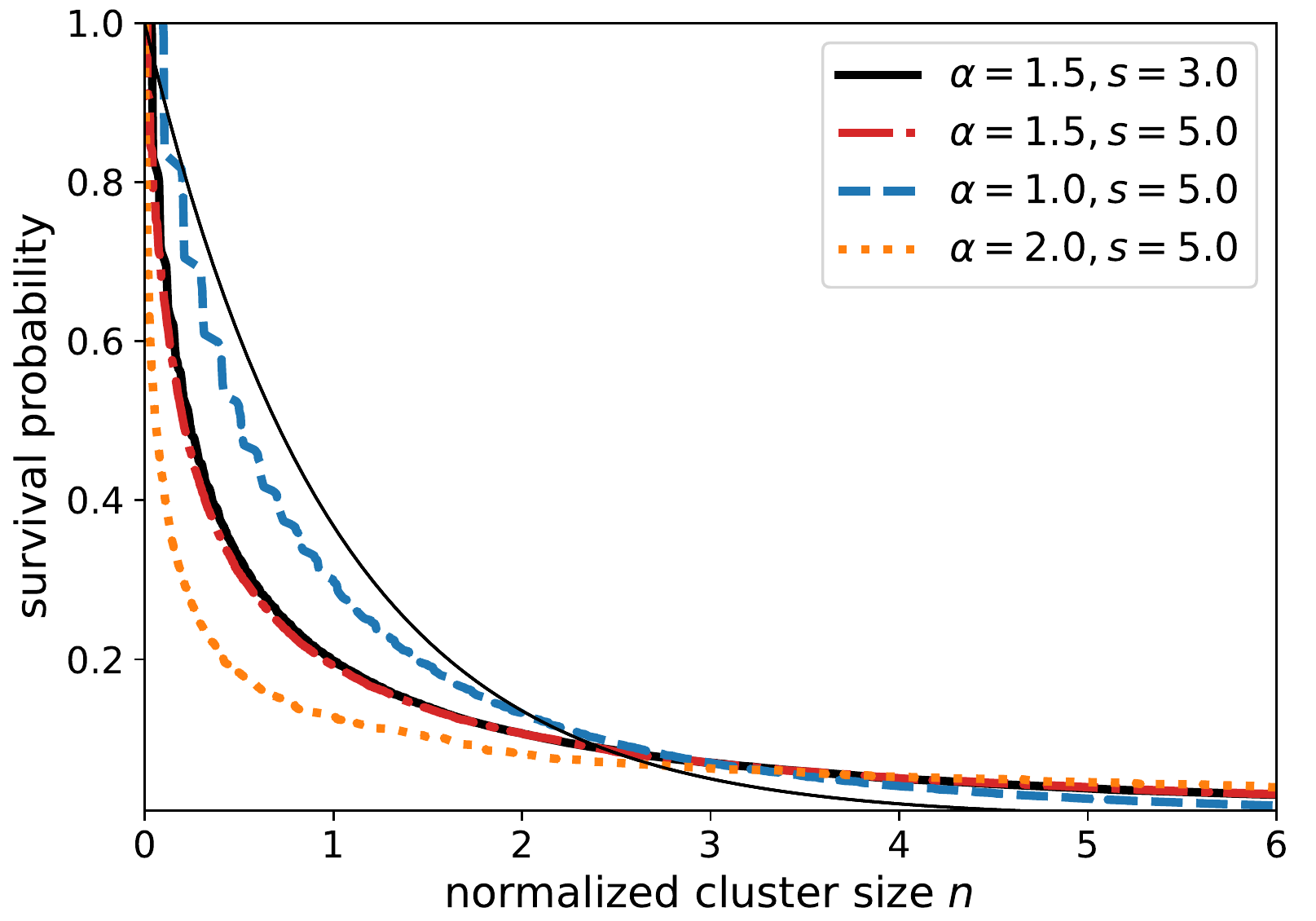}
  \caption{\label{fig:clusSize} Survival probability of
    the normalized cluster size for some LGGs in dimension 2. The first 2 curves
    (thick solid black
    and dotted-dashed red)
    have the same Levy index but different scales. The black one is
    barely noticeable since both lines superimpose.
    The following two (dashed blue and dotted orange) show the effect of varying $\alpha$
    within the LGG boundaries. 
    The scale used here was 5 but any other value
    would have given the same result. The thin black line shows the $e^{-n}$ 
    function.}

\end{figure}

We show in \Fig{clusSize} the measured survival probability of $n$ 
for Levy graphs for different indices and scales. 
The distributions are slightly milder than an exponential one 
and can be modeled by 
\begin{align}
\label{eq:nfit}
p(\geq n)\propto\exp(-\beta ~n^\gamma)  
\end{align}

with $\beta\simeq 2$ and $\gamma\simeq0.4$.

\begin{figure}
  \includegraphics[width=8.16cm]{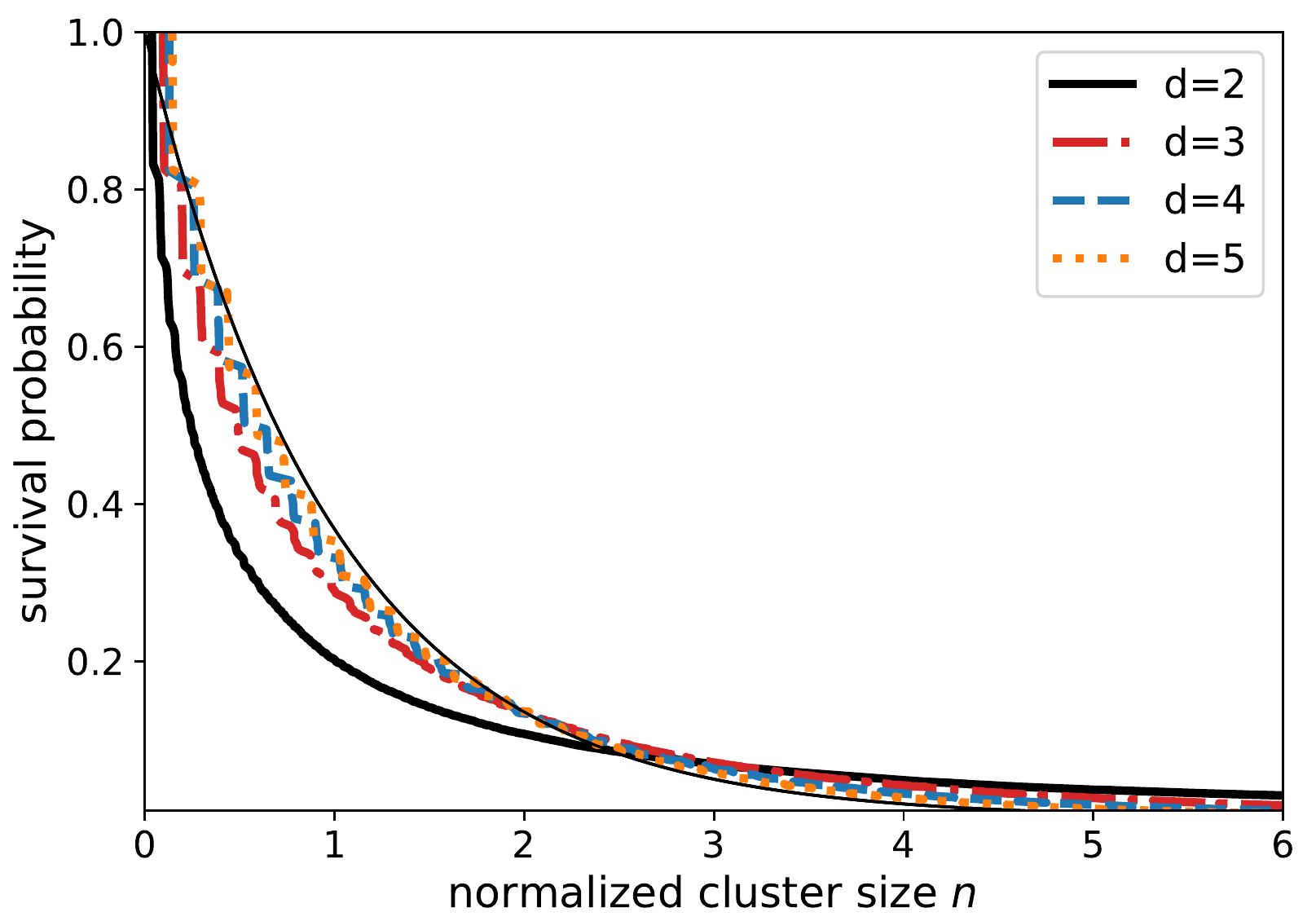}
  \caption{\label{fig:sizedim} Survival probability of
    the (normalized) clusters size when increasing the dimensionality
    $d$ of the space for $\alpha=1.5$. They converge to $e^{-n}$
    shown as the think black line.
    }
\end{figure}

To understand the origin of this shape we consider again the case of a large
dimension and show in \Fig{sizedim} the survival probability of $n$ when the 
dimension increases. The distribution becomes closer and closer to an
exponential type and seem to converge to $e^{-n}$.
In high dimensions, neglecting the return-probability, a 
cluster of size $N_i$ is formed from several small
steps and stops when a jump exceeds the scale
$s$, which happens with probability $p=\tfrac{1}{s^\alpha}$ (\refeq{surv}). 
Since the steps are independent, the distribution of the number
of points in the cluster is a geometric one:
\begin{align}
  p(N_i)&=(1-p)p^{N_i}\\
  & =\dfrac{1}{s^\alpha}\left(1-\dfrac{1}{s^\alpha}\right)^{N_i}.
\end{align}
We have seen that in this space $\Nclus/N=1/s^\alpha$, and  by
the change of variable $n=\tfrac{\Nclus}{N} N_i$
\begin{align}
  p(n)=\left(1-\dfrac{1}{s^\alpha}\right)^{s^\alpha n},
\end{align}
which, in the region we explore ($s^\alpha\gg 1$), converges indeed to $e^{-n}$.

\begin{figure}
  \includegraphics[width=8.16cm]{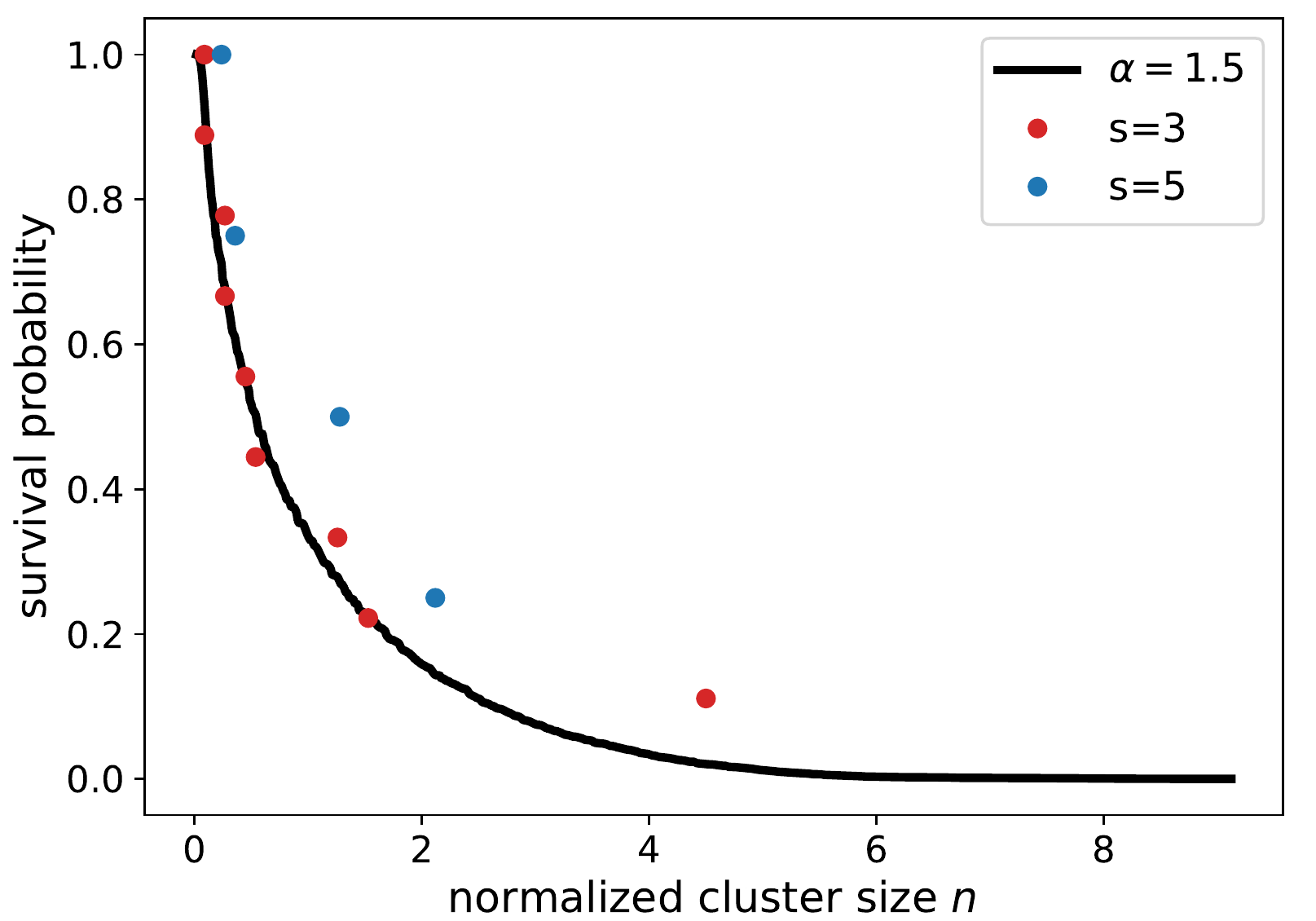}
  \caption{Normalized cluster size for $\alpha=1.5$ ($N=100$).
    Theoretical curve obtained from simulations in black and realizations observed on \Fig{lgg} for scales
    $s=3$ in red and 5 in blue.}
  \label{fig:csize_ex}
\end{figure}

But the most remarkable feature of the \Fig{clusSize} distributions 
is that they \textit{do not depend on the scale}. As an illustration, we
consider the graphs shown in \Fig{lgg}. For $s=3,5$ there are respectively
$\Nclus=9$ and 4 clusters and the normalized sizes are
\begin{subequations}
\begin{align}
\label{eq:nex}
  n(s=3)&=\dfrac{9}{100}(1,1,3,3,5,6,14,17,50)\\
  n(s=5)&=\dfrac{4}{100} (6,9,32,53)
\end{align}
\end{subequations}
If we rank those numbers and plot them on the theoretical curve
for $\alpha=1.5$, 
we see in \Fig{csize_ex} that they are both realizations of the \textit{same}
distribution, up to the noise due to the small statistics used for the illustration.
Results on a larger statistics is precisely what is shown in \Fig{clusSize}.

This statistical invariance comes from the self-similar nature of
the Levy flight meaning that the same complexity of the process is contained at any scale.
By building the LGG we capture this behavior into the
graph. A set of connected components at some given scale is equivalent to any other one
built at a different scale.
We have thus transferred the fractal geometry of the Levy points to the graph.

This allows to make a connection to more  abstract graphs, i.e. those
without a metric (as social networks). 
From their set of  connected components, we can test immediately
whether the
normalized sizes follow one of the \Fig{clusSize} distributions or not. If
not, they are incompatible with a LGG. If yes, we can associate a
potential $\alpha$ value, and from the fraction of clusters $\Nclus/N$ (\Fig{Nclus}), attribute a scale.
Further studies then need to be performed to test the topology of the
clusters in order to check if the graph could originate from a Levy
process.
The detailed clusters characterization is outside the scope of this paper
and we only illustrate it in the following on the mean degree.

\subsection{Clusters mean degree}
\label{sec:clusprop}

Although the full set of connected components provides an equivalent description of the graph
at any scale, a single cluster does not represent the entire graph.

Let us call $\kmean_i$ the mean degree of cluster $i$
\begin{align}
\kmean_i=\dfrac{1}{N_i} \sum_{j=1}^{N_i} k_j  .
\end{align}
The average degree of the graph can then be written
\begin{align}
\label{eq:meanki}
  \kmean&=\frac{1}{N}\sum_i N_i~\kmean_i=\frac{1}{\Nclus}\sum_in_i~\kmean_i,
\end{align}
by introducing the normalized cluster sizes $n_i$ (\refeq{normsize}).

\begin{figure}
  \includegraphics[width=8.16cm]{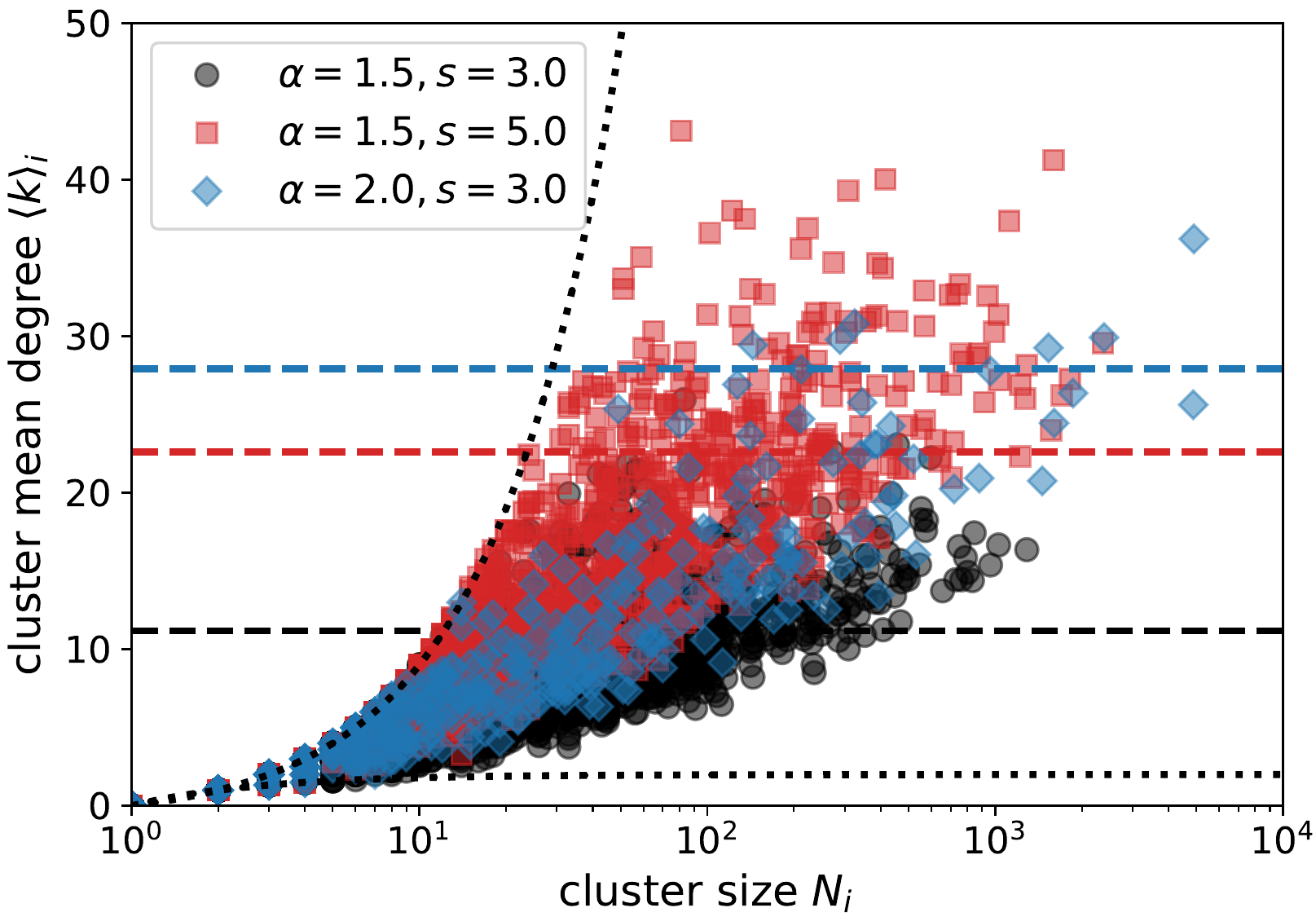}
  \caption{\label{fig:deg_clus} Mean degree of LGG clusters
    according to their
    size. Each point corresponds to one cluster in a $N=10^5$
    simulation.  The horizontal dashed lines show the graph's average degree.
    The dotted lines show the limits discussed in the text
    (\refeq{bounds}). 
    When $\alpha$ or $s$ increases 
    larger clusters may form for a fixed $N$ size run.}
\end{figure}

This expression captures the main dependence
on the LGG parameters since we have seen that  $\kmean\simeq
s^\alpha$  and $\Nclus\simeq 1/s^\alpha$. Accordingly, the sum should essentially not depend
on $s$ and $\alpha$. This is shown in \Fig{deg_clus} where the
distributions of the clusters mean degree vs. their size
are similar for different parameters of the LGG.

To understand the global shape, one must remember that the  distribution of
$n$ is peaked towards low values (\refeq{nfit}), 
so we expect many small size components. However the mean degree of a connected graph is constrained, especially
for low sizes.
For a cluster of size $N_i$ the
smallest degree is achieved with a path ($E_i=N_i-1$ edges) and
the largest one with a complete graph ($E_i=\tfrac{1}{2}N_i(N_i-1)$). From
$E_i=\kmean_i\tfrac{N}{2}$, the bounds on any cluster are therefore
\begin{align}
\label{eq:bounds}
  2\frac{N_i-1}{N_i} \leq~ \kmean_i ~\leq N_i-1,
\end{align}
corresponding to the gray areas in \Fig{deg_clus}.

These bounds are very constraining for low size clusters which are the most
numerous ones in LGGs. Then in order to maintain the graph's 
average degree verifying \refeq{meanki}, larger (rare) clusters must have
large degrees as observed in \Fig{deg_clus}. The important point here
is that the mean-degree is independent of $N$, so that \Fig{deg_clus}
is universal. Running with a higher $N$ value, one would (possibly) get a
few larger connected components which would add a few points on the right part
of the plot, but the main shape would remain unchanged.

Then each cluster plays a role in obtaining the correct graph's 
mean degree and a single one cannot be considered as a representation of the whole.

\section{Random walk graphs}
\label{sec:rw}

The new idea explored in this work is to build a
geometric graph on top of a random walk process. We may then ask what is
specific to Levy flights, which are very particular processes
with infinite variance steps.
We thus compare our results with a geometric graph built on top of a
standard random walk (SRW), i.e. with normally distributed increments of
variance $\sigma^2$.

We first consider the average degree for which we derive an
analytical formula in dimension 2 in the Appendix \ref{app:srw}:
\begin{align}
\label{eq:exact}
  \kmean&=2\sum_{k=1}^{N}\left(1-\dfrac{k}{N}\right)\left(1-e^{-\tfrac{s^2}{2k}}\right),
\end{align}
where the scale is defined here as $s\equiv\dfrac{R}{\sigma}$.

\begin{figure}
  \includegraphics[width=8.16cm]{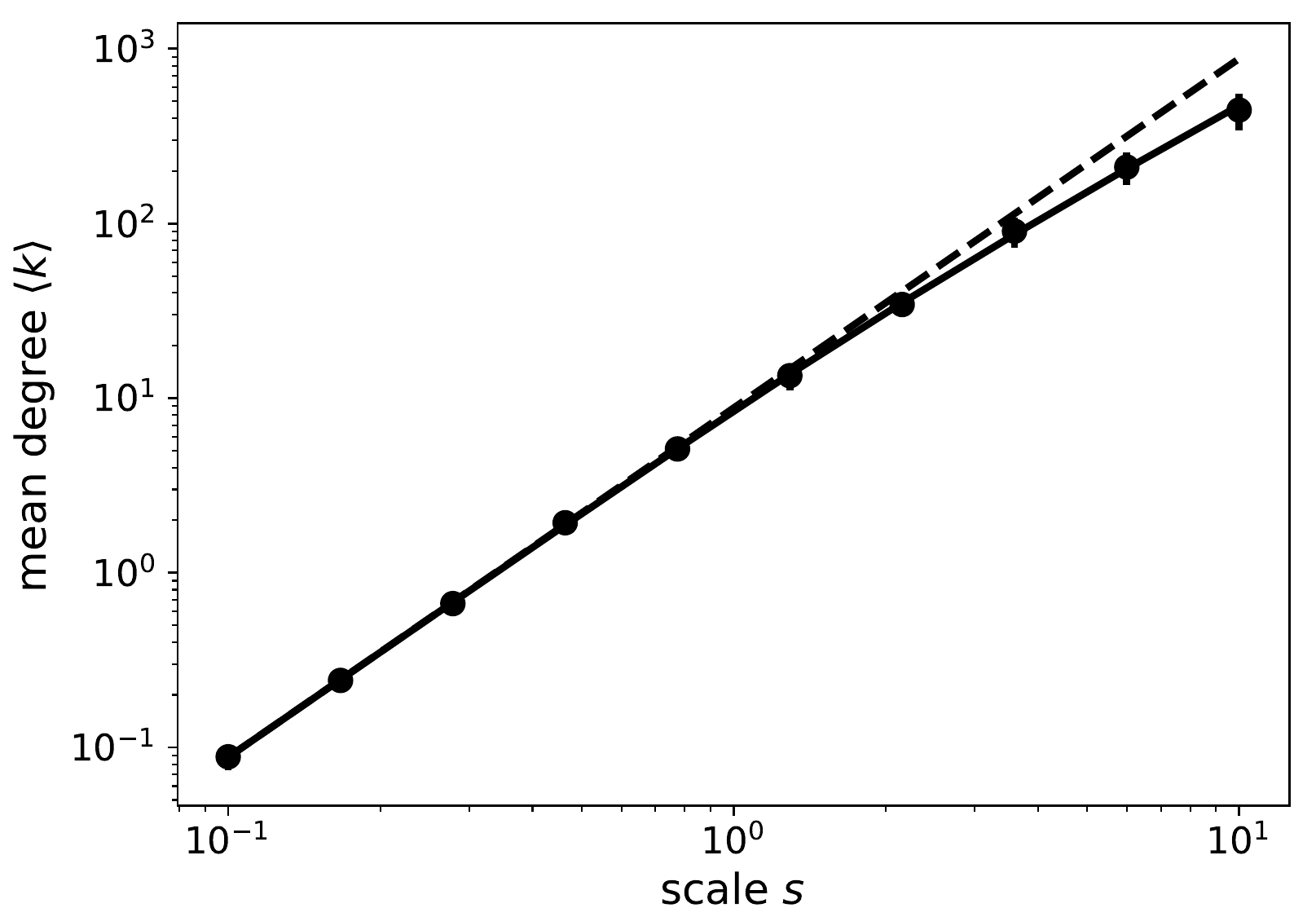}
  \caption{\label{fig:meandegRW} Mean degree for the standard random
    walk geometric graph
    depending on the scale cut $s=\dfrac{R}{\sigma}$ . The
    points with error bars show the 
    outcome of simulations ($N=10000,\Nsim=100$). The full line shows
    the exact analytical computation \refeq{exact}, and the dashed one
    is the \refeq{approx} quadratic approximation valid for $s\lesssim1$.
    }
\end{figure}

For $s\lesssim 1$ the argument of the exponential is small, so that
\begin{align}
\label{eq:approx}
  \kmean\simeq s^2 \sum_{k=2}^N\dfrac{1}{k},
\end{align}
which reveals a quadratic nature but only at low scales. 
Although formally diverging, the mean degree depends weakly on
  $N$ in practical cases (the sum being 8.8 for $N=10^4$ and 13.3 for the
$N=10^6$ case). 
We confront these calculations to simulations in \Fig{meandegRW} showing a perfect agreement.

As for the case of LGG, for which we had $\kmean\propto s^{\alpha_D}$
with $\alpha_D\lesssim1.5$ (\Fig{coeffs}), the
mean degree for SRWs looks approximately like a power-law (with $\alpha_D=2$).
But there is an important difference. While
for LGG the formula breaks down at \textit{low} scales (\Fig{meandeg}), for SRW it
breaks down at \textit{large} ones (\Fig{meandegRW}).

Another similarity comes from the degree distribution. We have checked
that for SRWs it is still well described by the $\Gamma$ distribution
(\sect{meandeg}). 
Then, using \refeq{exact} we also have an analytical description.

\begin{figure}
  \includegraphics[width=8.16cm]{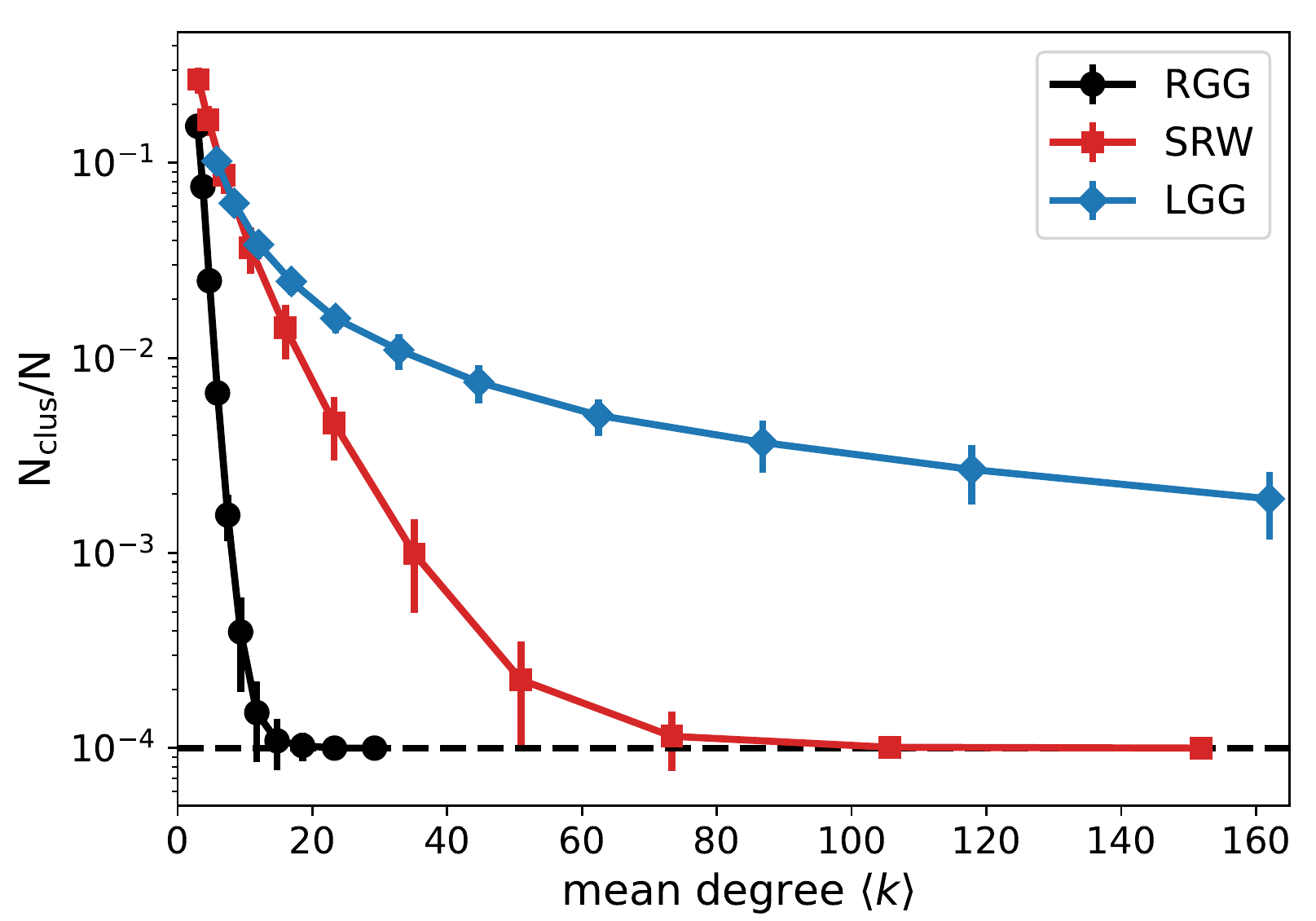}
  \caption{\label{fig:NclusRW}
    (a) Fractional number of clusters (for $N=10^4$) as a function of
    the mean degree for Random Geometric Graphs (RGG), Standard Random
    Walk ones (SRW) and Levy graphs (LGG,$\alpha=1.5$). The dashed line 
    indicates the \Nclus=1 case, i.e when there is a single giant component.
    }
\end{figure}

The main difference comes from the clusters.
We measure in \Fig{NclusRW} the fraction of clusters when increasing
the scale, or equivalently the mean degree, and added for reference the RGG case.
The SRW graph converges to a single cluster (the giant component) for
a connectivity about 10 times larger ($\simeq 50$) than for
the RGG. This corresponds to a scale around $s_c=2$ (see \Fig{meandegRW}) which
is the moment when the mean degree starts to deviate from a pure power-law.

For LGG, the power-law behavior stays exact and no giant component
ever appears when increasing the scale \footnote{Although technically one could
  imagine setting the scale to a huge 
number above the radius of the graph, it cannot be defined
\textit{a priori} since the maximal extent of a Levy graph
is unpredictable.}. 
This is not only due to the fact that the process is inhomogeneous (which 
can increase the threshold as in \citeg{Wang:2009} but not suppress the
transition), but to the fact that the point density goes to zero when increasing
the geometric cutoff $R$, since $\rho(R)=\tfrac{\bar N(<R)}{\pi R^2}\propto1/R^{2-\alpha_D}$ 
with $\alpha_D\lesssim1.5$ (\sect{meandeg}).
The set of points is asymptotically \textit{empty}:
a randomly-placed small volume contains typically no points, which
prevents the appearance of the giant component when increasing the radius.

In statistical physics language, the system never undergoes
a geometrical phase transition, as in percolation. This type of transition describes the
emergence of an ordered phase characterized by giant components:
highly connected clusters with sizes of the same order of magnitude as
$N$, i.e., macroscopic structures. At the critical point (or region),
though, clusters with various sizes coexist producing large
fluctuations in cluster statistics as can be noticed for RGG and SRW in \Fig{NclusRW}
slightly below the critical connectivity.  Traditional
random graphs represented here by SRW and RGG can only portray
critical behavior in a limited range. In the case of SRW, the typical power-law behavior 
holds up to scales $s_c\lesssim2$, indicating that beyond that point a
different theory and approximations must be employed to describe the
system. In contrast, for LGG  the scale invariance remains intact
and the same theory can be used, regardless of the scale used to
investigate the problem.

\section*{Conclusion}

We have investigated the properties of geometric graphs built on top
of random walk processes and in particular on Levy flights and found
the following:
\begin{itemize}
\item the mean degree is mostly independent of the graph's size,
\item it scales as a power-law of the geometric cut, $\kmean \propto R^\alpha$ 
  where $\alpha$ is the Levy index ( and is equal to $2$ for a standard
  random-walk graph but only for scales below $\sim2\sigma$),
\item the degree follows a $\Gamma$ distribution and has thus an exponential tail.
\end{itemize}
These are generic features of all (isotropic) random-walk graphs
since from the Generalized Central Limit Theorem, any process will either have a
finite variance and converge to a standard (Gaussian) walk, either
converge to a stable distribution with Levy-type tails.

We have thus found a simple way to construct a random geometric graph
with an exponential tail, i.e. broader than the standard Poisson
(Gaussian) one.

When considering the connected components (clusters) differences appear
between standard random walk graphs (i.e with finite variance steps)
and Levy-flight graphs (with infinite variance steps). The former show 
a critical connectivity much larger than for random geometric graphs. But the latter
show \textit{no critical transition at all}. For the Levy graph, a giant component
never forms, whatever the scale is.

For Levy graphs the number of clusters scales as an inverse power of
the scale.
By multiplying it by their size, one obtains a normalized
cluster size that is scale-invariant, i.e. that does not depend on the
geometric cutoff used to build the graph. 
Thus the set of clusters \textit{at any scale} is equivalent,  
which may be viewed as a generalization of the self-similar
nature of the Levy  flight from points to graphs.

This invariance can allow to make the connection to 
non-metric graphs by considering  only the size of their clusters. 
If the survival probability falls typically as $e^{-\beta n^\gamma}$
with $\beta\in[2,3]$ and $\gamma\in[0.3,1]$, one may associate a
potential Levy index, and from the fraction of clusters, a scale. 
To check further whether a graph could originate from a Levy process
or not, one needs to study the structure of its clusters. 
We have focused on degree distributions but several other topological
descriptors exist \cite{Newman:2003b}. We have found for instance that
the clustering coefficients 
(that is related to the density of triangles) is large (around
0.7); the average path lengths (shortest number of steps
between two vertices) scales as $N^{1/d}$ and is therefore not
compatible with a ``small-world'' network \cite{AB:2002}.
These two aspects come from the the local nature
of the geometric cutoff that favors triangles and forbids the appearance
of long shortcuts. 

Levy graphs may find application in several areas.

On the theoretical side, they reveal an intriguing feature:
although they exhibit several power-law dependencies that are
characteristic of critical regions \cite{Stanley:1999,AB:2002}, they
actually never experience a  transition. Could it be that they are  
\textit{always} in a critical state? They could then serve as a
prototype for studying systems close to a critical point.

Our second finding is that systems without an intrinsic scale
but analyzed at a given scale show a very characteristic distribution
of their cluster sizes. This may find applications in community
detection. Many methods exist to identify communities in a graph but the
scale at which to search for them is unclear  \cite{Fortunato:2010}.
Then by running a single algorithm, one can check the cluster
characteristics and possibly attribute a Levy index. 

Are there some data to which we can confront our model to?
To this aim we need to turn on to scale-free systems that are common
in biology \cite{Mora:2011}, as in the flock of birds \cite{Cavagna:2010}.
More generally, the analysis and modeling of \textit{collective behaviors} may
be an interesting target, as in the self-organization of pedestrian
crowds that show some Levy-walk strategies\cite{Murakami:2019}. 
But the most direct application could be to the modeling of face-to-face interactions.
Some high-quality data that record the time individuals meet in various
environments are available \footnote{\safeurl{www.sociopatterns.org}}
and are best analyzed with aggregated graphs \cite{Holme:2012}. 
Several important aspects,
as the distribution of contact duration, are well described by graphs
built on random-walks \cite{Starnini:2013}. Biased random-walks can also capture the
appearance of recurrent communities \cite{Flores:2018}.
It is then natural to explore whether Levy walks may be beneficial to this
field since the  appearance of communities (clusters) lies at the very 
heart of Levy graphs.

\begin{acknowledgments}
  We acknowledge the use of the \textsf{graph-tool} package
  \safeurl{https://graph-tool.skewed.de} for all graph-related computations.
\end{acknowledgments}

\appendix

\section{Conditional probability distribution of a 2D Levy process}
\label{app:xi}

We detail in this appendix the computation of the conditional
distribution for a Levy process in the plane. We follow
closely \cite{Peebles:1980} by adapting it to dimension 2 (since it
was performed in dimension 3 ) enriching the demonstration and
quantifying approximations being made.

We start from a point of the process.
From \refeq{surv} the probability distribution of the next
displacement in the plane is
\begin{align}
f_1(\bm{r})=
  \begin{cases}
    \dfrac{\alpha}{2\pi}\dfrac{r_0^\alpha}{r^{\alpha+2}}  & \rm{for}~ r \geq r_0 \\
    0 & \rm{otherwise}.
  \end{cases}
\end{align}
The process being isotropic, its generating function (Fourier
transform) only depends on the mode modulus $k$. Integrating over the angles
\begin{align}
  \psi_1(k)&=\int f_1(\bm{r}) e^{i \bm{k}\cdot\bm{r}} d^2\bm{r} \\
  &=\alpha r_0^\alpha \int_{r_0}^\infty
             \dfrac{J_0(kr)}{r^{\alpha+1}} dr
\end{align}
where we used \cite{GR} (8.411-7)
\begin{align}
  \label{eq:bess}
  \int_{0}^{2\pi} e^{\pm i z \cos\phi}d\phi&=2\pi J_0(z),
\end{align}
$J_0$ being a Bessel function of first type.

Integrating by parts
\begin{align}
\label{eq:psi1}
  \psi_1(k)&=J_0(kr_0)-k r_0^\alpha\int_{r_0}^\infty \dfrac{J_1(kr)}{r^{\alpha}} dr
\end{align}
using \cite{GR} (6.511-7) $[J_0(kr)]^\prime=-kJ_1(kr)$ 

We are interested in the $r\gg r_0$ case so that $kr_0\ll1$ and
\begin{align}
  J_0(kr_0)\simeq1-\tfrac{(kr_0)^2}{4}
\end{align}

For $0<\alpha<2$ the integral gets most of its contribution from the
$r>r_0$ tail so that we can use \cite{GR} (6.561-14)
\begin{align}
\label{eq:besspow}
\int_{0}^\infty x^\mu J_m(ax) dx &= 2^\mu
  \tfrac{\Gamma(1/2+m/2+\mu/2)}{\Gamma(1/2+m/2-\mu/2)}~a^{-\mu-1}\nn \\\text{for}&-m-1<\mu<1/2
\end{align}
to obtain


\begin{align}
\label{eq:psi11}
  \psi_1(k)&\simeq 1-I_\alpha \left(kr_0\right)^\alpha,\\
  \text{with}~& I_\alpha=\tfrac{\Gamma(1-\alpha/2)}{2^\alpha \Gamma(1+\alpha/2)}\nn.
\end{align}
One recognizes the asymptotic characteristic function of stable
distributions ($ \exp(-\sigma^\alpha k^\alpha)$) corresponding to the
heavy tail of the Pareto-Levy distribution \cite{Chechkin:2008}.

The generating function for the $n^{th}$ displacement is the product of the
individual functions
\begin{align}
  \psi_n(k)=\psi_1^n(k),
\end{align}
and the probability distribution its inverse Fourier transform
\begin{align}
  f_n(\bm{r})&=\tfrac{1}{(2\pi)^2}\int \psi_n(k) e^{-i
               \bm{k}\cdot\bm{r}} dk.
\end{align}
Considering \textit{any} number of steps
\begin{align}
  f(\bm{r})&=\sum_n f_n=\tfrac{1}{(2\pi)^2} \int \sum_n \psi_1^n(k) e^{-i
               \bm{k}\cdot\bm{r}} d^2\bm{k} \nn \\
           &=\tfrac{1}{(2\pi)^2} \int \left[1-\psi_1(k)\right]^{-1} e^{-i
               \bm{k}\cdot\bm{r}} d^2\bm{k} \nn \\
  &=  \dfrac{I_\alpha^{-1} r_0^{-\alpha}}{(2\pi)^2}\int k^{-\alpha} e^{-i
               \bm{k}\cdot\bm{r}} d^2\bm{k} \nn \\
&= \dfrac{I_\alpha^{-1} r_0^{-\alpha}}{2\pi}\int_0^\infty
                                                k^{1-\alpha}J_0(kr) dk,
\end{align}
where we use again \refeq{bess} when integrating over the angles.
From \refeq{besspow}
\begin{align}
\int_0^\infty k^{1-\alpha}J_0(kr) dk=K_\alpha r^{\alpha-2}~\text{with}~K_\alpha=\tfrac{\Gamma(1-\alpha/2)}{2^{\alpha-1}\Gamma(\alpha/2)},
\end{align}
and we finally find that for $\alpha<2$ and $r\gg r_0$
\begin{align}
  f(\bm{r})&=\dfrac{C}{r^{2-\alpha}},\quad C=\tfrac{\Gamma(1+\alpha/2)}{\pi \Gamma(\alpha/2)} r_0^{-\alpha}.
\end{align}

For $\alpha\ge2$, the integral \refeq{besspow} diverges in the $r_0\to0$ limit.
In fact it now gets most of its contribution from low $r$ values i.e. around $r_0$
where $J_1(kr)\simeq kr/2$. With this crude approximation
\begin{align}
\label{eq:psib}
  \psi_1(k)&\simeq 1-\dfrac{3}{4}\left(kr_0\right)^2.
\end{align}

This is the leading order of a small Gaussian displacement.  
Its inverse-Fourier transform is then also a Gaussian and one recovers
(roughly) a standard random walk.

\begin{figure}
  \centering
  \includegraphics[width=8.16cm]{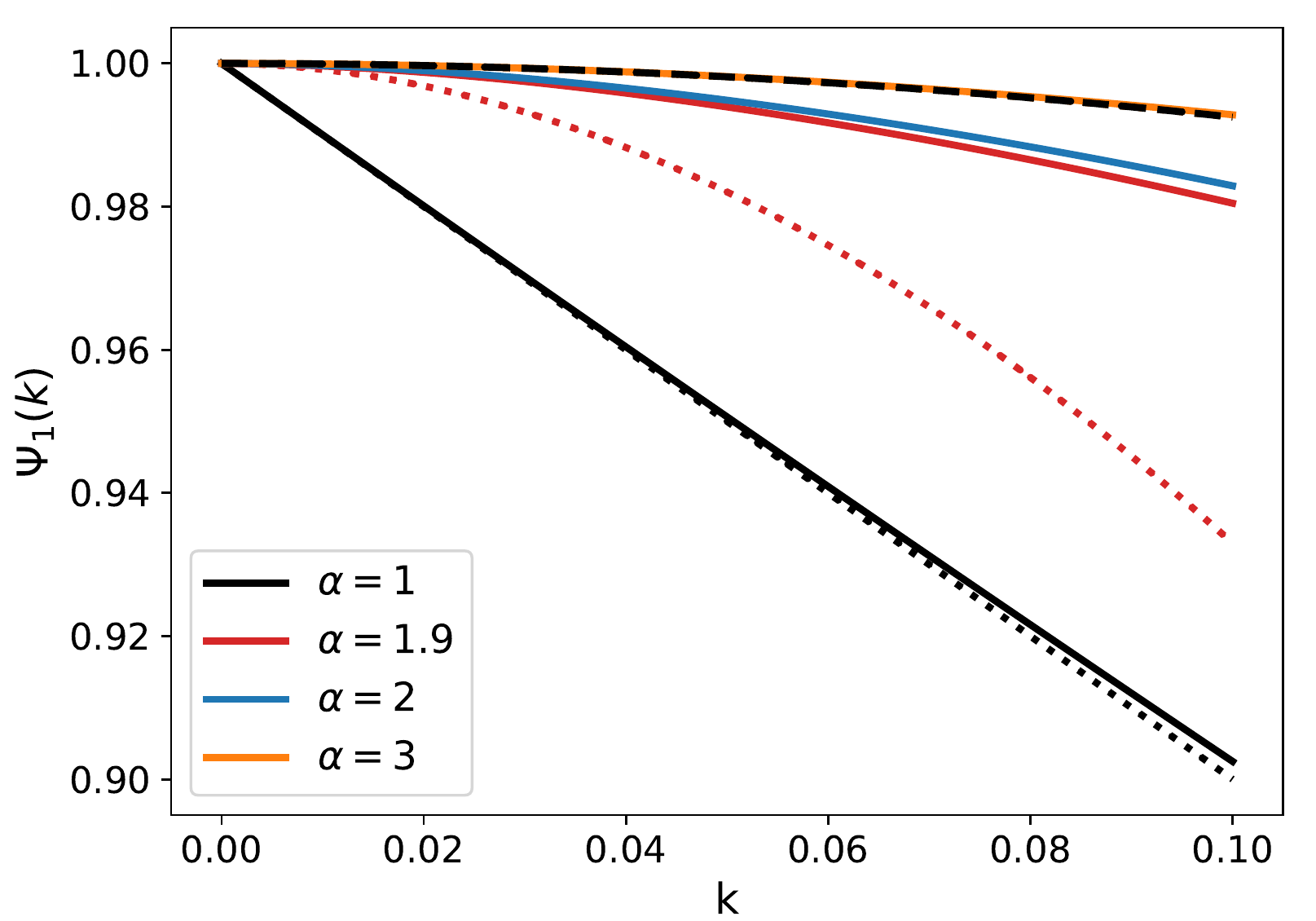}
  \caption{\label{fig:psi1}Test of the approximations made in deriving the
    conditional probability function of a 2D Levy process. The full
    lines show the exact numerical computation of the single-step characteristic function
    \refeq{psi1}. Dotted lines shows the power-law
    approximation \refeq{psi11} used in the $\alpha<2$ case. The black
    dashed line shows the quadratic approximation \refeq{psib} used
    for all $\alpha\ge2$ indices. We use $r_0=1$ and consider
    the $kr_0\ll1$ region.}
\end{figure}

We can (and should) question the rather strong simplifications that
were made to the \refeq{psi1} integral in both the $\alpha<2$ and $\alpha\ge2$
regimes. With $r_0=1$, we compare in \Fig{psi1} the exact
value of $\psi_1(k)$ from \refeq{psi1} computing numerically the
integral, to the derived approximations which are \refeq{psi11} for
$\alpha<2$ and \refeq{psib} for $\alpha\geq2$. 
The approximation is excellent for $\alpha=1$ but gets worse when
approaching 2. For $\alpha=2$ the quadratic approximation is not
yet reached and becomes satisfactory only around $\alpha=3$.

\section{Mean degree of Standard Random Walk graphs}
\label{app:srw}
In a standard (Gaussian) random-walk process, the coordinates of the
increments follow a normal distribution of variance $\sigma^2$, that we
note in dimension 2, $x_k,y_k\sim\N{0,\sigma^2}$.
The coordinates of the $i^{th}$ point in the walk, as the sum of
independent normal variables, then follow $X_i,Y_i\sim \N{0,i\sigma^2}$.
Let us focus on a point at index $t$ and compute the distance of any other
point at index $i$ to it
\begin{align}
\label{eq:ray}
r_{ti}=\sqrt{(X_t-X_i)^2+(Y_t-Y_i)^2} .
\end{align}
since $X_t-X_i=\sum_{k=1}^tx_k-\sum_{k=1}^i x_k=\sum_{k=i+1}^t x_k$
assuming $i<t$, without loss of generality
\begin{align}
X_t-X_i\sim \N{0,\abs{t-i}\sigma^2}.  
\end{align}
The same holds independently for $Y_t-Y_i$, so that \refeq{ray} represent the
distance between two normally distributed independent variables, 
each of variance $\abs{t-i}\sigma^2$. It then follows a Rayleigh
distribution of cumulative function
\begin{align}
  P_{t,i}(<R)=1-e^{-\tfrac{R^2}{2\abs{t-i}\sigma^2}}.
\end{align}
Let us now consider $\bar N_t$ the mean number of points within some
distance $R$ of point $t$. Each point has a probability $P_{t,i}(<R)$
to be in the vicinity of $t$, so that
\begin{align}
  \bar N_t&= \sum_{i=1}^N P_{t,i}(<R)
\end{align}
where, for $i=t$, we set $P_{i,i}=0$ so as to only count neighbors.
The mean degree of the geometric graph with a $R$ distance cutoff is
obtained by averaging $N_t$ over all the $t$ points:
\begin{align}
  \kmean&=\dfrac{1}{N}\sum_{t=1}^N\bar N_t\nn \\
  &=\dfrac{1}{N}\sum_{t=1}^N \sum_{i=1}^N  \left(1-e^{-\tfrac{s^2}{2\abs{t-i}}}\right)
\end{align}
where we introduce the relevant scale $s\equiv\dfrac{R}{\sigma}$.

We may simplify the formula by noticing that
$P_{t,i}(<R)$ is a circulant matrix symmetric around the $P_{t,t}=0$
diagonal and that the double sum represents
the sum of all its elements. Then by counting the elements along the diagonals
\begin{align}
  N\kmean=2(N-1)(1-e^{-\tfrac{s^2}{2}})+2(N-2)(1-e^{-\tfrac{s^2}{4}})+\cdots
\end{align}
and finally
\begin{align}
  \kmean&=2\sum_{k=1}^{N}\left(1-\dfrac{k}{N}\right)\left(1-e^{-\tfrac{s^2}{2k}}\right).
\end{align}


\bibliographystyle{unsrtnat}

\bibliography{levy}

\begin{thebibliography}{34}
\providecommand{\natexlab}[1]{#1}
\providecommand{\url}[1]{\texttt{#1}}
\expandafter\ifx\csname urlstyle\endcsname\relax
  \providecommand{\doi}[1]{doi: #1}\else
  \providecommand{\doi}{doi: \begingroup \urlstyle{rm}\Url}\fi

\bibitem[Erd\"os and R\'enyi(1959)]{ER:1959}
P.~Erd\"os and A.~R\'enyi.
\newblock On random graphs.
\newblock \emph{Publicationes Mathematicae}, 6:\penalty0 290--297, 1959.

\bibitem[Erd\"os and R\'enyi(1960)]{ER:1960}
P.~Erd\"os and A.~R\'enyi.
\newblock On evolution of random graphs.
\newblock \emph{Publ. Math. Inst. Hung. Acad. Sci.}, 6:\penalty0 17--60, 1960.

\bibitem[Erd\"os and R\'enyi(1961)]{ER:1961}
P.~Erd\"os and A.~R\'enyi.
\newblock On the strength of connectedness of random graphs.
\newblock \emph{Acta. Math. Sci. Hung}, 12:\penalty0 261--267, 1961.

\bibitem[Albert and Barab\'asi(2002)]{AB:2002}
R\'eka Albert and Albert-L\'aszl\'o Barab\'asi.
\newblock Statistical mechanics of complex networks.
\newblock \emph{Rev. Mod. Phys.}, 74:\penalty0 47--97, 2002.

\bibitem[Gilbert(1961)]{Gilbert:1961}
E.~N. Gilbert.
\newblock Random plane networks.
\newblock \emph{Journal of the Society for Industrial and Applied Mathematics},
  9\penalty0 (4):\penalty0 533--543, December 1961.
\newblock \doi{10.1137/0109045}.
\newblock URL \url{https://doi.org/10.1137/0109045}.

\bibitem[Dall and Christensen(2002)]{Dall:2002}
Jesper Dall and Michael Christensen.
\newblock Random geometric graphs.
\newblock \emph{Phys. Rev. E}, 66\penalty0 (1):\penalty0 016121, July 2002.
\newblock ISSN 1063-651X, 1095-3787.
\newblock \doi{10.1103/PhysRevE.66.016121}.
\newblock URL \url{https://link.aps.org/doi/10.1103/PhysRevE.66.016121}.

\bibitem[Newman(2003)]{Newman:2003b}
M.~E.~J. Newman.
\newblock The {Structure} and {Function} of {Complex} {Networks}.
\newblock \emph{SIAM Rev.}, 45\penalty0 (2):\penalty0 167--256, January 2003.
\newblock ISSN 0036-1445, 1095-7200.
\newblock \doi{10.1137/S003614450342480}.
\newblock URL \url{http://epubs.siam.org/doi/10.1137/S003614450342480}.

\bibitem[Barthélemy(2011)]{Barthelemy:2011}
Marc Barthélemy.
\newblock Spatial networks.
\newblock \emph{Physics Reports}, 499\penalty0 (1-3):\penalty0 1--101, February
  2011.
\newblock ISSN 03701573.
\newblock \doi{10.1016/j.physrep.2010.11.002}.
\newblock URL
  \url{https://linkinghub.elsevier.com/retrieve/pii/S037015731000308X}.

\bibitem[Herrmann et~al.(2003)Herrmann, Barthelemy, and Provero]{Herrmann:2003}
Carl Herrmann, Marc Barthelemy, and Paolo Provero.
\newblock Connectivity {Distribution} of {Spatial} {Networks}.
\newblock \emph{Phys. Rev. E}, 68\penalty0 (2):\penalty0 026128, August 2003.
\newblock ISSN 1063-651X, 1095-3787.
\newblock \doi{10.1103/PhysRevE.68.026128}.
\newblock URL \url{http://arxiv.org/abs/cond-mat/0302544}.
\newblock arXiv: cond-mat/0302544.

\bibitem[Krioukov et~al.(2010)Krioukov, Papadopoulos, Kitsak, Vahdat, and
  Bogu\~n\'a]{Krioukov:2010}
Dmitri Krioukov, Fragkiskos Papadopoulos, Maksim Kitsak, Amin Vahdat, and
  Mari\'an Bogu\~n\'a.
\newblock Hyperbolic geometry of complex networks.
\newblock \emph{Phys. Rev. E}, 82:\penalty0 036106, Sep 2010.
\newblock \doi{10.1103/PhysRevE.82.036106}.
\newblock URL \url{https://link.aps.org/doi/10.1103/PhysRevE.82.036106}.

\bibitem[{Barab{\'a}si} and {Albert}(1999)]{BA:1999}
Albert-L{\'a}szl{\'o} {Barab{\'a}si} and R{\'e}ka {Albert}.
\newblock {Emergence of Scaling in Random Networks}.
\newblock \emph{Science}, 286\penalty0 (5439):\penalty0 509--512, October 1999.
\newblock \doi{10.1126/science.286.5439.509}.

\bibitem[{Van Kampen}(2007)]{Kampen:2007}
N.G. {Van Kampen}.
\newblock \emph{Stochastic Processes in Physics and Chemistry (Third Edition)}.
\newblock North-Holland Personal Library. Elsevier, Amsterdam, third edition
  edition, 2007.
\newblock \doi{https://doi.org/10.1016/B978-044452965-7/50011-8}.
\newblock URL
  \url{https://www.sciencedirect.com/science/article/pii/B9780444529657500118}.

\bibitem[{Hughes}(1995)]{Hugues:1995}
B.D. {Hughes}.
\newblock \emph{Random Walks and Random Environments. Volume 1: Random Walks}.
\newblock Clarendon Press, Oxford, 1995.
\newblock URL
  \url{https://global.oup.com/academic/product/random-walks-and-random-environments-9780198537885?cc=fr&lang=en&}.

\bibitem[Durrett(2010)]{Durrett:2010}
R.~Durrett.
\newblock \emph{Probability: {Theory} and examples}.
\newblock Cambridge series in statistical and probabilistic mathematics.
  Cambridge University Press, 2010.
\newblock ISBN 978-1-139-49113-6.
\newblock URL \url{https://books.google.fr/books?id=evbGTPhuvSoC}.

\bibitem[Mandelbrot(1975)]{Mandelbrot:1975}
Benoit Mandelbrot.
\newblock "sur un modèle décomposable d'univers hiérarchisé: déduction des
  corrélations galactiques sur la sphère céleste.".
\newblock \emph{Comptes Rendus (Paris)}, 280A:\penalty0 1551--1554, 1975.
\newblock URL
  \url{https://users.math.yale.edu/mandelbrot/web_pdfs/comptes_rendus_79.pdf}.

\bibitem[{Mandelbrot}(1983)]{Mandelbrot:1983}
Benoit {Mandelbrot}.
\newblock \emph{{The Fractal Geometry of Nature}}.
\newblock Freeman, San Francisco, 1983.

\bibitem[Chechkin et~al.(2006)Chechkin, Gonchar, Klafter, and
  Metzler]{Chechkin:2006}
Aleksei~V. Chechkin, Vsevolod~Y. Gonchar, Joseph Klafter, and Ralf Metzler.
\newblock Fundamentals of l{\'{e}}vy flight processes.
\newblock In \emph{Fractals, Diffusion, and Relaxation in Disordered Complex
  Systems}, pages 439--496. John Wiley {\&} Sons, Inc., June 2006.
\newblock \doi{10.1002/0470037148.ch9}.
\newblock URL \url{https://doi.org/10.1002/0470037148.ch9}.

\bibitem[Cox and Isham(1980)]{Cox:1980}
D.R. Cox and V.~Isham.
\newblock \emph{Point Processes}.
\newblock Chapman \& Hall/CRC Monographs on Statistics \& Applied Probability.
  Taylor \& Francis, 1980.
\newblock ISBN 9780412219108.
\newblock URL \url{https://books.google.fr/books?id=KWF2xY6s3PoC}.

\bibitem[Chechkin et~al.(2008)Chechkin, Metzler, Klafter, and
  Gonchar]{Chechkin:2008}
Alexei~V. Chechkin, Ralf Metzler, Joseph Klafter, and Vsevolod~Yu. Gonchar.
\newblock In Rainer Klages, Gnter Radons, and Igor~M. Sokolov, editors,
  \emph{Introduction to the Theory of Levy Flights}, pages 129--162. Wiley-VCH
  Verlag GmbH \& Co. KGaA, Weinheim, Germany, July 2008.
\newblock ISBN 978-3-527-62297-9 978-3-527-40722-4.
\newblock \doi{10.1002/9783527622979.ch5}.
\newblock URL
  \url{https://onlinelibrary.wiley.com/doi/10.1002/9783527622979.ch5}.

\bibitem[Newman(2005)]{Newman:2005}
M.~E.~J. Newman.
\newblock Power laws, {Pareto} distributions and {Zipf}'s law.
\newblock \emph{Contemporary Physics}, 46\penalty0 (5):\penalty0 323--351,
  September 2005.
\newblock ISSN 0010-7514, 1366-5812.
\newblock \doi{10.1080/00107510500052444}.
\newblock URL \url{http://arxiv.org/abs/cond-mat/0412004}.
\newblock arXiv: cond-mat/0412004.

\bibitem[Montroll(1956)]{Montroll:1956}
Elliot~W. Montroll.
\newblock Random walks in multidimensional spaces, especially on periodic
  lattices.
\newblock \emph{Journal of the Society for Industrial and Applied Mathematics},
  4\penalty0 (4):\penalty0 241--260, 1956.
\newblock \doi{10.1137/0104014}.

\bibitem[Note1()]{Note1}
Note1.
\newblock Although technically one could imagine setting the scale to a huge
  number above the radius of the graph, it cannot be defined \protect \textit
  {a priori} since the maximal extent of a Levy graph is unpredictable.

\bibitem[Wang and González(2009)]{Wang:2009}
Pu~Wang and Marta~C. González.
\newblock Understanding spatial connectivity of individuals with non-uniform
  population density.
\newblock \emph{Philosophical Transactions of the Royal Society A:
  Mathematical, Physical and Engineering Sciences}, 367\penalty0
  (1901):\penalty0 3321--3329, August 2009.
\newblock ISSN 1364-503X, 1471-2962.
\newblock \doi{10.1098/rsta.2009.0089}.
\newblock URL
  \url{https://royalsocietypublishing.org/doi/10.1098/rsta.2009.0089}.

\bibitem[Stanley(1999)]{Stanley:1999}
H~Eugene Stanley.
\newblock Scaling, universality, and renormalization: {Three} pillars of modern
  critical phenomena.
\newblock \emph{Rev. Mod. Phys.}, 71\penalty0 (2):\penalty0 9, 1999.

\bibitem[Fortunato(2010)]{Fortunato:2010}
Santo Fortunato.
\newblock Community detection in graphs.
\newblock \emph{Physics Reports}, 486\penalty0 (3-5):\penalty0 75--174,
  February 2010.
\newblock ISSN 03701573.
\newblock \doi{10.1016/j.physrep.2009.11.002}.
\newblock URL
  \url{https://linkinghub.elsevier.com/retrieve/pii/S0370157309002841}.

\bibitem[Mora and Bialek(2011)]{Mora:2011}
Thierry Mora and William Bialek.
\newblock Are {{Biological Systems Poised}} at {{Criticality}}?
\newblock \emph{Journal of Statistical Physics}, 144\penalty0 (2):\penalty0
  268--302, July 2011.
\newblock ISSN 0022-4715, 1572-9613.
\newblock \doi{10.1007/s10955-011-0229-4}.

\bibitem[Cavagna et~al.(2010)Cavagna, Cimarelli, Giardina, Parisi, Santagati,
  Stefanini, and Viale]{Cavagna:2010}
Andrea Cavagna, Alessio Cimarelli, Irene Giardina, Giorgio Parisi, Raffaele
  Santagati, Fabio Stefanini, and Massimiliano Viale.
\newblock Scale-free correlations in starling flocks.
\newblock \emph{Proceedings of the National Academy of Sciences}, 107\penalty0
  (26):\penalty0 11865--11870, June 2010.
\newblock ISSN 0027-8424, 1091-6490.
\newblock \doi{10.1073/pnas.1005766107}.

\bibitem[Murakami et~al.(2019)Murakami, Feliciani, and
  Nishinari]{Murakami:2019}
Hisashi Murakami, Claudio Feliciani, and Katsuhiro Nishinari.
\newblock L\'evy walk process in self-organization of pedestrian crowds.
\newblock \emph{Journal of The Royal Society Interface}, 16\penalty0
  (153):\penalty0 20180939, April 2019.
\newblock ISSN 1742-5689, 1742-5662.
\newblock \doi{10.1098/rsif.2018.0939}.

\bibitem[Note2()]{Note2}
Note2.
\newblock \protect \href {www.sociopatterns.org}{www.sociopatterns.org}.

\bibitem[Holme and Saram{\"a}ki(2012)]{Holme:2012}
Petter Holme and Jari Saram{\"a}ki.
\newblock Temporal {{Networks}}.
\newblock \emph{Physics Reports}, 519\penalty0 (3):\penalty0 97--125, October
  2012.
\newblock ISSN 03701573.
\newblock \doi{10.1016/j.physrep.2012.03.001}.

\bibitem[Starnini et~al.(2013)Starnini, Baronchelli, and
  {Pastor-Satorras}]{Starnini:2013}
Michele Starnini, Andrea Baronchelli, and Romualdo {Pastor-Satorras}.
\newblock Modeling {{Human Dynamics}} of {{Face-to-Face Interaction Networks}}.
\newblock \emph{Physical Review Letters}, 110\penalty0 (16):\penalty0 168701,
  April 2013.
\newblock ISSN 0031-9007, 1079-7114.
\newblock \doi{10.1103/PhysRevLett.110.168701}.

\bibitem[Flores and Papadopoulos(2018)]{Flores:2018}
Marco Antonio~Rodr{\'i}guez Flores and Fragkiskos Papadopoulos.
\newblock Similarity {{Forces}} and {{Recurrent Components}} in {{Human
  Face-to-Face Interaction Networks}}.
\newblock \emph{Physical Review Letters}, 121\penalty0 (25):\penalty0 258301,
  December 2018.
\newblock ISSN 0031-9007, 1079-7114.
\newblock \doi{10.1103/PhysRevLett.121.258301}.

\bibitem[{Peebles}(1980)]{Peebles:1980}
P.~J.~E. {Peebles}.
\newblock \emph{{The large-scale structure of the universe}}, chapter III.62.
\newblock Princeton University Press, 1980.

\bibitem[{Gradshteyn} and {Ryzhik}(2007)]{GR}
I.~S. {Gradshteyn} and I.~M. {Ryzhik}.
\newblock \emph{{Table of Integrals, Series, and Products}}.
\newblock Academic Press, 2007.

\end{thebibliography}


\end{document}